# Material-Property-Field-based Deep Neural Network in Hopfield Framework

*Yanxiao Hu [1, #], Ye Sheng [1, #], Caichao Ye [1,3,4,*], Wenxing Qian [1], Xiaoxin Xu [1, 2], Yabei Wu [1], Jiong Yang [2], William A. Goddard III [3], Wenqing Zhang [1, 4, *]*

*[1] State Key Laboratory of Quantum Functional Materials & Department of Materials Science and Engineering, Southern University of Science and Technology, Shenzhen 518055, China.*

*[2] Materials Genome Institute, Shanghai Engineering Research Center for Integrated Circuits and Advanced Display Materials, Shanghai University, Shanghai 200444, China.*

*[3] Materials and Process Simulation Center (MSC), California Institute of Technology, Pasadena 91125, California, United States*

*[4] Guangdong Provincial Key Laboratory of Computational Science and Material Design & Institute of Innovative Materials, Academy for Advanced Interdisciplinary Studies, Southern University of Science and Technology, Shenzhen 518055, Guangdong, China.*

*[#] Authors contributed equally.*

** Corresponding Authors: W. Zhang: zhangwq@sustech.edu.cn; C. Ye: yecc@sustech.edu.cn*

## ABSTRACT

Current deep neural networks (DNNs) used in materials modeling often lack explicit physical structure and clear analytical formulations tailored to material systems, which can limit their interpretability. In this work, we integrate Material Property Fields (MPF) with the Hopfield network architecture and propose an analytically structured DNN framework named mPFDNN. MPF provides a unified framework that represents physical properties of materials as an analytical field built upon pairwise interactions, rigorously respecting fundamental symmetries, while also enabling a physically legitimate decomposition of property distributions at the atomic level. Although the Hopfield model was originally developed for Ising-like systems, we show that its dynamical evolution strategy can be naturally extended to the MPF framework. By reformulating nonlinear interatomic interactions as "hidden neurons", MPF can be extended into a deep yet analytically tractable DNN architecture that progressively captures an increasingly connected interaction landscape. This framework also provides a unified perspective that connects linear expansions and nonlinear DNN architectures within a common interaction-based formulation. Extensive validation across diverse systems, including inorganic crystals, organic molecules, and aqueous solutions, and across multiple target properties, shows that mPFDNN achieves competitive predictive accuracy while offering a physically motivated framework for structure-property mapping in chemistry, physics, and materials science.

## 1. INTRODUCTION

A central objective of physics, chemistry and materials science is to establish robust, quantitatively predictive relationships that link composition and structure to properties. However, this pursuit is inherently limited by the intrinsic complexity of the structure-property relationship, which resides in a high-dimensional space of strong nonlinearity exceeding the capacity of human intuition. In recent years, the integration of artificial intelligence into materials science [1-8], particularly through deep neural networks (DNNs) [9-12], marked a turning point. By stacking multiple nonlinear hidden layers, DNNs acquire the theoretical capacity to approximate any continuous functions[13], thereby enabling the description of high-order interatomic interactions. Simply by applying this off-the-shelf technique, current material-oriented DNNs have achieved significant success that traditional paradigms could hardly attain. This impressive performance, however, is often accompanied by limited interpretability and unclear physical transparency. Because explicit mathematical structure and physically motivated inductive biases are often limited, the generalization behavior and physical interpretability of such models can remain difficult to assess. In this context, the integration of bottom-up data-driven modeling with top-down physics-based modeling has emerged as a crucial frontier in the application of artificial intelligence to science[14-16].

According to the first Hohenberg-Kohn theorem [17], the ground-state properties of physical and chemical systems are fundamentally determined by its atomic coordinates. This determinative relationship can be formulated and modeled through effective interatomic interactions. Notably, the modeling of these interactions need not be treated as a wholly uncharacterized "black box"; instead, it can be guided by well-established physical priors and expressed in chemically meaningful forms. It is a fundamental insight that pairwise interatomic interactions are strongly dependent on interatomic distance, while many-body effects must be captured via directional features such as bond angles and dihedral angles. Consequently, the interaction space serves as an intermediary that bridges atomic coordinates and macroscopic properties, thereby offering a unique language for the modeling of real systems.

Motivated by this insight, we proposed in our previous work [18] the concept of Material Property Fields (MPF). This is an analytical field representation inspired by wave functions in quantum many-body systems, providing a unified framework for physical property defined as a functional distribution in real space (e.g., total energy, stress, and dielectric constant). In this framework, pairwise interactions serve as fundamental building blocks, endowing MPF with a mathematical structure that inherently satisfies all required symmetries. Building on this foundation and incorporating global scaling laws, we further developed SUS$^2$-MLIP [18], an ultra-compact analytical machine-learning interatomic potential (MLIP) with a linear basis architecture. It delivers accuracy on par with advanced DNNs while using 2-3 orders of magnitude fewer parameters and achieving substantially higher simulation efficiency. This line of thought naturally raises the question of whether MPF can be translated into a DNN architecture, and what role interatomic interactions should play within such a network. Addressing this question may help improve the physical transparency of materials-oriented neural-network models.

To this end, we turn to the Hopfield network, which is built upon the interactions and collective dynamical evolution inherent to many-body systems. In the classical formulation [19] of Hopfield network, the pairwise interactions between neurons refer to the Ising model of magnetic spins. Leveraging the tendency of thermodynamic systems to evolve toward steady states, the Hopfield network achieves memory functionality by tailoring specific interaction rules that place important information at the bottom of the energy landscape. When a state (e.g., a noisy or incomplete memory) is presented to a trained model, dynamical evolution automatically drives it downhill to the nearest energy minimum, i.e. the stored memory pattern. This mechanism has been utilized to solve linear and nonlinear equations of mathematical physics [20]. By constructing a Hopfield network that the minima of its energy function correspond exactly to the solutions of the target problem, an approximate solution, when input as an initial state, can be automatically refined through dynamical evolution to converge to the exact solution. Recent studies [21] have further extended this paradigm by employing modern Hopfield networks as "predictor-corrector" models for forecasting the evolution of complex systems, where an approximate solution for the system's next state is first obtained via a crude model, and a Hopfield network then serves as a corrector to learn and rectify the prediction errors, yielding more accurate results. Drawing on this strategy, this work proposes mPFDNN, a Hopfield-inspired framework in which "interatomic interactions" are treated as "hidden neurons" that drive the recursive evolution of the system representation. Starting from a mean-field representation of MPF, it progressively recovers the modulating role of the chemical environment on interactions, and thereby spontaneously evolves into a fully connected interaction landscape.

The performance of mPFDNN was evaluated across diverse systems, including solids, molecules, and liquid solutions, where it achieved competitive accuracy with substantially fewer parameters in many cases. Notably, the strong predictive capability of mPFDNN for adsorption energies on high-entropy alloy catalysts highlights our model's robust extrapolation performance, even for complex and out-of-domain systems. By combining competitive predictive accuracy with an analytically motivated interaction-based formulation, mPFDNN provides a useful framework for the design of physically informed neural networks in materials modeling.

## 2. THEORETICAL FRAMEWORK

### 2.1 Theory of Material Property Field and MPF-based DNN framework

In density functional theory, the ground-state properties of a system are completely determined by the nuclear coordinates $\{\boldsymbol{x}\}$. In this context, all ground-state-dependent properties $P$ are also the function of $\{\boldsymbol{x}\}$, and it can also be decomposed into a sum of atomic contributions $\{P_I\}$ as:

$$P(\{\boldsymbol{x}\}) = \sum_{I=1}^{N} P_I\left(\{\boldsymbol{x}\}\right), \tag{1}$$

It is important to emphasize that $P_I$ depends on the coordinates of all atoms $\{\boldsymbol{x}\}$, which ensures that the decomposition fully captures arbitrary long-range interactions, including nucleus-nucleus electrostatic interactions. This decomposition is legitimized by the fact that the total electron density can itself be partitioned into atomic contributions, even if such partitions are not unique (e.g., Bader analysis, Hirshfeld analysis, and Mulliken population analysis). Naturally, $P_I$ must also satisfy fundamental symmetry constraints, including translational invariance, permutation invariance with respect to atomic indices, and rotational covariance corresponding to the specific property of interest.

In coordinate space, perturbations of atomic nuclei lead to alterations in the overall electron density distribution, consequently changing the properties of the system. This process inherently involves exchange-correlation effects in many-electron systems and thus intricate interatomic interactions, thereby establishing $P_I(\{\boldsymbol{x}\})$ a high-dimensional nonlinear function. As shown in **Figure 1a,** DNNs abstractly encapsulate this intricate physical relationship in the interconnections between feature neurons as Y = σ(Wx), where the weight matrix W encodes node correlations and the activation function σ fits nonlinear patterns. Yet this implicit formulation, while statistically powerful, leaves the true physical nature of $P_I$ obscured.

To render the underlying physics explicit and analytically tractable, we define $P_I$ in interaction space as the $P[\{\varphi_{Ij}\}]$. As illustrated in **Figure 1b**, the $\varphi_{Ij}(\boldsymbol{r_{Ij}}, v_I, v_J)$ represents the effective interaction of atom *j* on atom *I*, which is a function of the position vector $\boldsymbol{r_{Ij}} = \boldsymbol{x_j} - \boldsymbol{x_I}$, and the intrinsic states $v_I$ and $v_J$ of the atoms. $v_I$ is dominated by its elemental factors as nuclear charge, while being strongly modulated by the surrounding chemical environment. The function $\varphi_{Ij}$ inherently satisfies translational invariance, and consequently, so does the $P[\{\varphi_{Ij}\}]$. As swapping the index {*j*} should not alter the physical outcome, drawing inspiration from the physically-clear representation of wave functions in quantum many-body systems, we express it in a product form and name it as material property field (MPF) [18]:

$$P = \sum_I P_I = \sum_I \prod_{j \neq I} (\varphi_{Ij}(\boldsymbol{r_{Ij}}, v_I, v_j) + 1). \tag{2}$$

MPF provides a mathematical mapping from coordinate space to property space, mediated by effective interactions. This formulation is physically well grounded, drawing analogy to tensor network approaches in which high-order many-body correlations are represented as tensor products defined on the Hilbert spaces of individual subsystems, as seen in Slater determinants for many-electron systems and the order parameter expansion in Landau's theory [22]. The "+1" factor prevents the product from vanishing as any individual interaction term $\varphi_{Ij}$ approaches zero, consistent with the logic of the Mayer f-function [23] in statistical mechanics. Moreover, upon expansion, this construction naturally generates a hierarchical series spanning many-body contributions, as will be demonstrated in Eq. (7). Importantly, the resulting interaction terms are physically motivated, in contrast to general DNN formulations that rely on multiple nonlinear active layers.

To capture rotational covariance, $\varphi_{Ij}$ is expanded using a complete set of SO(3) basis functions, such as the spherical harmonics $Y_l^m$. This expansion is then expressed as a tensor product:

$$\varphi_{Ij} = v_I \otimes \vec{\mathfrak{R}}\left(\boldsymbol{r}_{Ij}\right) \otimes v_j, \tag{3}$$

where the tensor representation of $\vec{\mathfrak{R}}\left(\boldsymbol{r}_{Ij}\right) = \oplus_{lm} R_l\left(|r_{Ij}|\right) Y_l^m\left(\hat{r}_{Ij}\right)$ is used for physical completeness. The $R_l\left(|r_{Ij}|\right)$ is the radial weight function to be determined. In our previous work, $v_I$ and $v_J$ were treated as quantities determined solely by the elemental factor $Z$, which corresponds to a mean-field approximation. This simplification inevitably introduces systematic errors, because the state of an atom is not fixed by its elemental identity alone but is modulated by its local chemical environment and neighboring atoms. To account for these indirect environmental effects, we turn to the Hopfield network, a framework rooted in collective interactions and dynamical evolution in many-body systems. Because such systems naturally evolve toward steady states, Hopfield-type networks have been widely used to solve linear and nonlinear equations and to refine predictions in complex systems [20,21]. In this context, the mean-field approximation provides a physically motivated initial solution to MPF and a natural starting point for iterative refinement. We then show that, by taking MPF as the governing equation in a Hopfield-like interaction scheme, the modulation of $\{v_I\}$ by the chemical environment can be incorporated recursively, yielding an analytical network-based formulation of MPF.

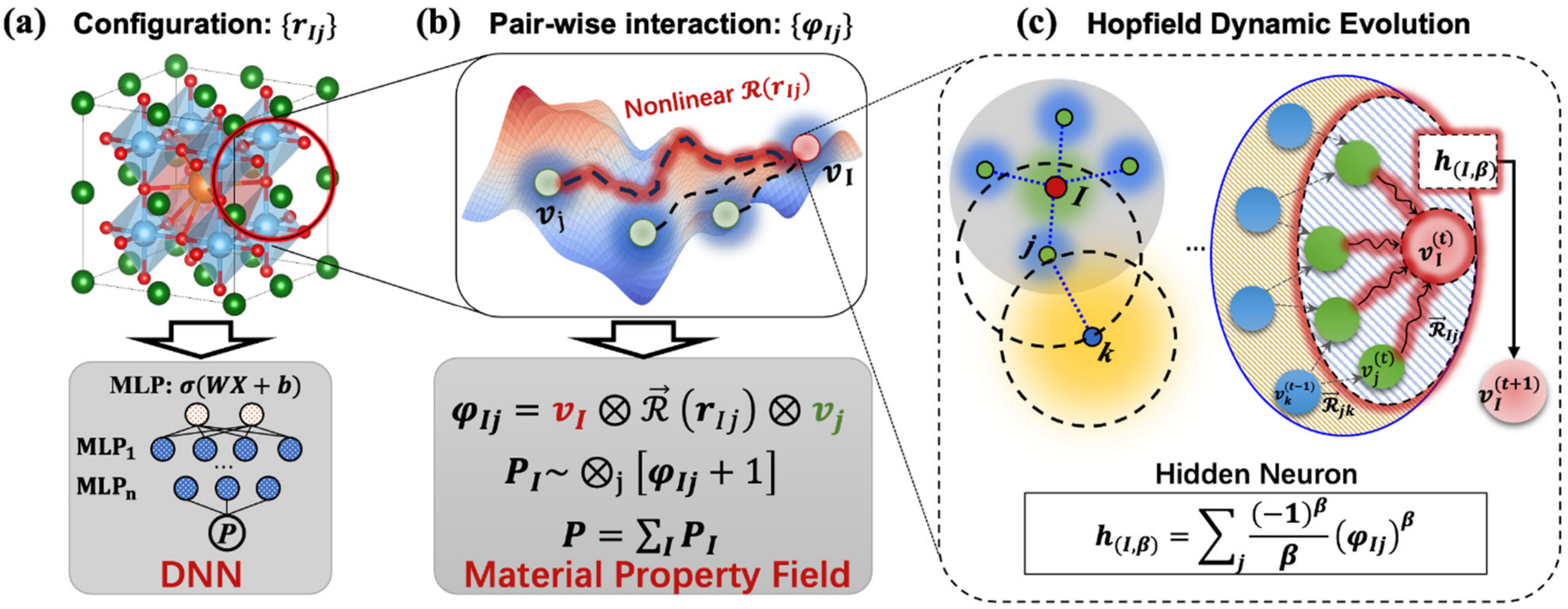

**Figure 1. (a)** Conventional deep neural networks (DNNs) that learn nonlinear structure–property relationships via empirical and trial-and-error active functions; **(b)** the Material Property Field (MPF) framework that constructs these relationships via analytically derived, physically complete tensor products; and **(c)** Hopfield dynamics that analytically integrate both local chemical environment effects and indirect long-range correlations through interaction-based hidden neurons.

In a Hopfield network, the system state is represented by a set of feature neurons $\{v_I\}$. In an Ising-like setting, for example, each $v_I$ takes a binary value of either $+1$ or $-1$. Interactions among these neurons drive the system dynamics and generate collective correlations between different components. In modern Hopfield frameworks with many-body couplings, such correlations are encoded implicitly through a set of hidden neurons $\{h_\mu\}$. The coupling between feature and hidden neurons then defines the effective energy landscape of the system. For a model containing $N_f$ feature neurons and $N_h$ hidden neurons, the "energy" is given by[24]:

$$E = \sum_{i}^{N_f} v_i g_i - L_v - L_h, \tag{4}$$

where $L_v(\{v_i\})$ and $L_h(\{h_\mu\})$ denote the Lagrangians for feature and hidden neurons, with $g_i = \frac{\partial L_v}{\partial v_i}$. Typically, $L_v = \frac{1}{2}\sum_I v_I^2$ and $L_h$ shapes the global interaction patterns and thus dictates the system's dynamical evolution.

Substituting the explicit form of the MPF into Eq. (4) with $L_h = \sum_I \prod_{j\neq I}(\varphi_{Ij} + 1)$ yields the energy function:

$$E = \frac{1}{2}\sum_{I} v_I^2 + \sum_{I} exp\,(\sum_{\beta\geq 1} h_{\mu=(I,\beta)}), \tag{5}$$

where the hidden neuron $h_\mu \equiv h_{(I,\beta)} = \sum_j \frac{(-1)^\beta}{\beta}\left(\varphi_{Ij}\right)^\beta$ aggregate all pairwise interactions centered on atom $I$, with $\beta$ implicitly covering the full $(l, m)$ expansion space and higher-order terms. A complete derivation is given in Section I of the SI. From Eq. (5) the dynamical evolution of the feature nodes $\{v_I\}$ simplifies to:

$$v_I^{(t+1)} = \sum_{\mu}^{N_h} \xi_{I\mu} \frac{\partial L_h^{(t)}}{\partial h_\mu^{(t)}} = \sum_{\mu}^{N_h} \xi_{I\mu}\, P_I^{(t)}, \tag{6}$$

where $\xi_{I\mu}$ is an adjustable Hopfield network parameter quantifying the contribution of hidden neuron $\mu$ to node $I$. By imposing the normalization condition $\sum_\mu^{N_h} \xi_{I\mu} = 1$, the dynamics reduce to a direct update rule of $v_I^{(t+1)} = P_I^{(t)}$, where the field at step $t$ recursively determines the atomic state at step $t+1$. As depicted in **Figure 1c**, beginning with the mean-field approximation $v_I^0 \equiv v_{Z_I}$, the model progressively incorporates environmental effects, first through local-neighbor interactions and then through more subtle long-range correlations, ultimately building a fully connected interaction landscape reminiscent of a wave function delocalized in space. Importantly, this formulation is derived analytically from the MPF framework, rather than introduced through empirical architectural trial and error.

### 2.2 Unification of Linear Approach with Scalable Nonlinear DNNs

Following the theoretical framework developed above, we construct mPFDNN, an analytically derived model that integrates linear basis with DNNs. A schematic of its architecture is provided in **Figure 2a**.

First, by expanding the product form in Eq. (2), ignoring constant terms, we obtain a body-ordered expansion:

$$P_I = \sum_j \varphi_{Ij} + \frac{1}{2}\left(\sum_j \varphi_{Ij}\right)^2 + \frac{1}{6}\left(\sum_j \varphi_{Ij}\right)^3 + \cdots \tag{7}$$

In theory, $P_I$ includes all high-order interactions. Introducing a tensor function $\boldsymbol{\mathcal{A}}_I = \sum_j \left[v_I \otimes \vec{\Re}\left(\boldsymbol{r}_{Ij}\right) \otimes v_j\right]$ allows the above expression to be rewritten more compactly as:

$$P_I =< \boldsymbol{\mathcal{P}}_I >= \sum_N \frac{1}{N!} \langle {\boldsymbol{\mathcal{A}}_I}^{\otimes N} \rangle \tag{8}$$

where $\boldsymbol{\mathcal{P}}_I$ is a tensor including the complete irreducible representation space of SO(3) group, and the $\langle\,\rangle$ indicates symmetry constraints such as rotational invariance, as required by specific properties. The $(N+1)$-body nonlinear interaction is represented as $v$ times direct product of pair function, i.e. the $\otimes^N$ in Eq. (8). Two examples of three-body and four-body are shown in **Figure 2b**.

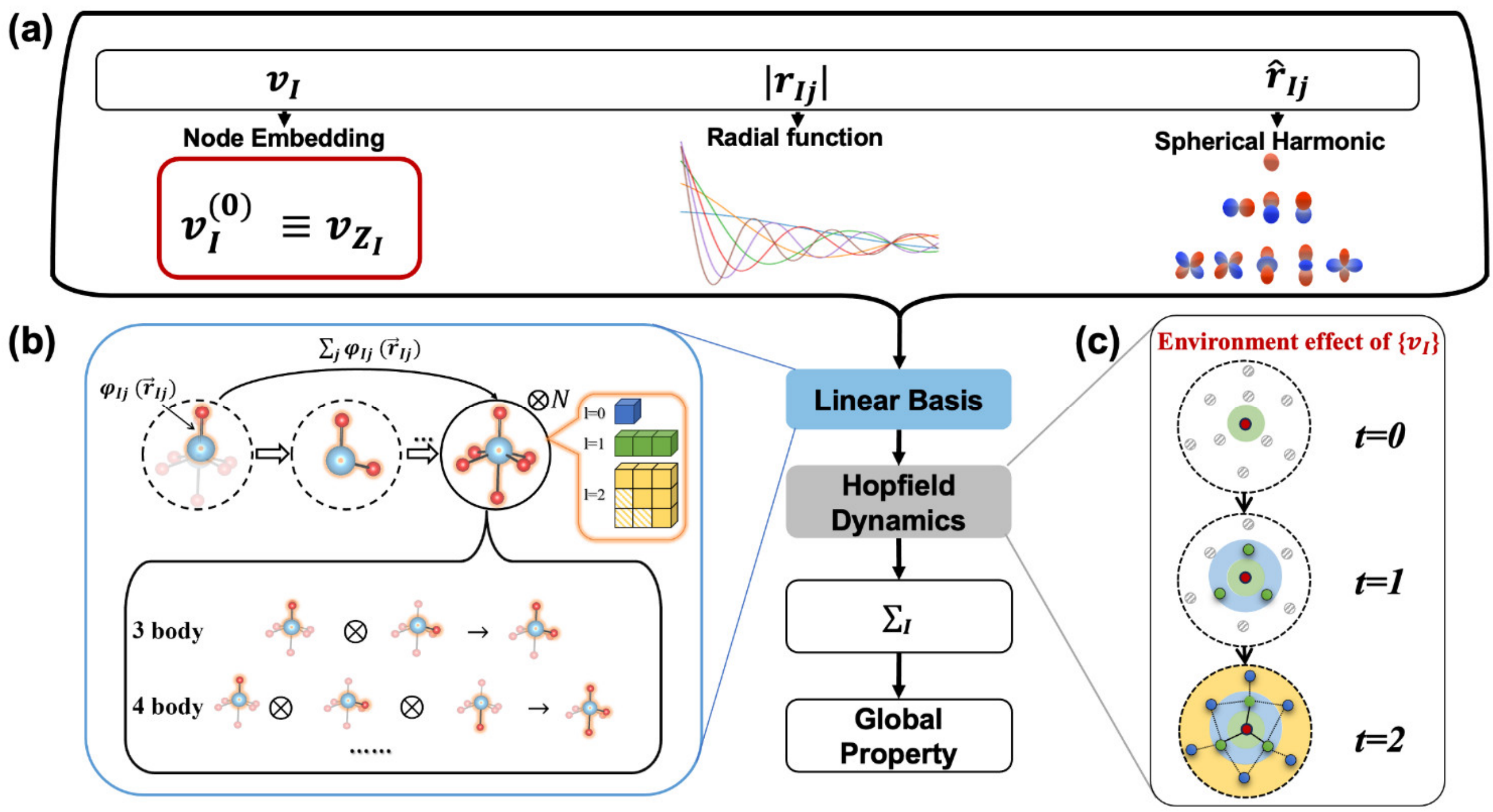

**Figure 2**. **(a)** End-to-end pipeline of mPFDNN. Atomic pairs serve as fundamental building blocks to encode atomic structures. **(b)** Starting from a body-ordered linear expansion under the mean-field approximation, **(c)** the framework evolves via Hopfield dynamics, capturing chemical environment correlations and yielding an analytical DNN architecture.

Through initially summing over neighboring atoms and subsequently applying the tensor product operation, Eq.(8) can obtain the aggregate of all body-ordered interactions while preventing a significant increase in computational complexity. Note that the tensor function $\boldsymbol{\mathcal{A}}_I \equiv \oplus_{lm} \boldsymbol{\mathcal{A}}_I^{lm}$, covers the full $\{l,m\}$ space. This allows us to expand Eq. (8) into a linear-like form, where Clebsch-Gordan coefficients $C_{l_1m_1\ldots l_nm_n}^{LM}$ are employed to select the irreducible representation subspaces $(L,M)$ consistent with rotational covariance:

$$
\begin{aligned}
P_I &= \sum_{n=0}^{N} \sum_{\substack{\alpha \\ \in (\boldsymbol{l_1m_1}\ldots \boldsymbol{l_nm_n}) \to \boldsymbol{LM}}} c_\alpha \, C_{l_1m_1\ldots l_nm_n}^{LM} \mathcal{A}_I^{l_1m_1} \ldots \mathcal{A}_I^{l_nm_n} \\
&= \tilde{B}_I^0 + \sum_{n=1}^{N} \sum_{\substack{\alpha \\ \in (\boldsymbol{l_1m_1}\ldots \boldsymbol{l_nm_n}) \to \boldsymbol{LM}}} c_\alpha \tilde{B}_I^\alpha ,
\end{aligned}
\tag{9}
$$

where $\alpha$ indexes the chosen reduction path of the representation space, and the coefficients $c_\alpha$ are parameters to be determined. As $LM=00$, it naturally degenerates into a scalar field. The self-interaction of $\tilde{B}_I^0 = \frac{1}{2} < v_I \otimes v_I >_{LM}$ corresponds the first term of Eq. (5), which in conventional DNN models is typically treated as a constant bias dependent solely on element type. It is worth noting that although Eq. (9) takes the form of a linear expansion, it inherently captures the nonlinear physics of many-body interactions, which laid the groundwork for the development of SUS$^2$-MLIP in our prior work [18].

To generalize this framework while maintaining physical transparency embodied in Eq. (9), we incorporate Hopfield dynamics via substituting the tensor form of Eq. (8) into the iterative relation of Eq. (6). This yields a recursive dynamics that naturally gives rise to deep network architectures in an analytically tractable form:

$$
\begin{cases}
\boldsymbol{\mathcal{A}}_I^{(\mathbf{t})} = \sum_j \boldsymbol{v}_{\boldsymbol{I}}^{(\boldsymbol{t})} \otimes \vec{\mathfrak{R}}\left(\boldsymbol{r}_{Ij}\right) \otimes \boldsymbol{v}_{\boldsymbol{j}}^{(\boldsymbol{t})} \\
v_I^{(t+1)} = \sum_{v} \left[\boldsymbol{\mathcal{A}}_I^{(t)}\right]^{\otimes v}
\end{cases}
\tag{10}
$$

As illustrated in **Figure 2c**, at t=0, the effective interatomic interactions $\{\varphi_{Ij}\}$ are determined solely by the geometric and elemental information of each atomic pair. After one recursive step, $v_j$ becomes aware of and modulated by its surrounding chemical environment. Specifically, the recursive approach systematically captures indirect correlations, such as those arising from multi-step interaction pathways (e.g., $k \to j \to I$) that are inherently absent in the mean-field approximation. Theoretically, as $t \to \infty$, the model converges to a fully connected interaction landscape.

Note that $v_j$ has been tensorized during the iterative process. As the present work focuses on scalar properties, we restrict $v_j^{(t)}$ to the irreducible representation subspace with $L' = 0$ (denoted as $v_j^{(t),L'=0}$) when updating $A_I^{(t)}$. The more general formulation is provided in Section I of the SI. Besides, in our implementation, the scalar radial functions $R_l(r_{Ij})$ are expanded to form a $k$-channel tensor $R_l^{(k)}(r_{Ij})$. This expansion, which has no impact on the analytical derivations, is equivalent to integrating $k$ models to enhance expressive capability.

In summary, the mPFDNN framework unifies linear basis expansions and DNNs within a coherent, analytically derived architecture. Within this framework, the physical transparency of the former and the representational capacity of the latter coexist synergistically, thereby offering a principled alternative to purely data-driven DNNs for materials modeling.

## 3. RESULTS

### 3.1 Comprehensive Benchmarking Demonstrates Broad Applicability and Accuracy

**Table 1 Summary of benchmark datasets.**

| Category | Dataset | Sample Number | Property |
|---|---|---|---|
| Crystal | Jarvis [25] | 75,908 | Formation Energy, $E$<br>Total Energy, *Total E*<br>Voigt bulk, $K_v$<br>Shear modulus, $G_v$<br>Dielectric constant, $\epsilon$ |
| | MPtraj (M3GNet) [26] | 187,687 | Energy/Force/Stress |
| Molecule | QM9 [25] | 133,885 | Isotropic polarizability, $\alpha$<br>Electronic spatial extent, $r^2$<br>Dipole moment, $\mu$<br>Internal energy at 0K, $U_0$ |
| | Drug [27] | 1.4 Million | Energy/Force |
| | OC2M [28] | 2 Million | Energy/Force |
| Metal | Alloy [29] | 72,722 | Energy/Force |
| Aqueous solutions | Water & Salt-water [30] | 33,819 | Energy/Force |
| HEA | IrPdPtRhRu [30] | 16,772 | Energy/Force |

To rigorously evaluate the universality and accuracy of mPFDNN, we curated a diverse set of benchmark datasets spanning inorganic crystals, organic molecules, and catalytic systems [30]. As summarized in **Table 1**, these datasets cover a broad range of important target properties, including standard potential-energy-surface quantities such as energies and atomic forces, as well as electronic properties derived from first-principles calculations, such as polarizabilities, dielectric constants, and dipole moments. The elemental coverage spans nearly the entire periodic table (**Figure 3a**), providing a stringent testbed for assessing the model's transferability.

Across these benchmarks, mPFDNN delivers either the best or highly competitive performance relative to state-of-the-art models. On the Mptraj dataset for inorganic crystals and the OC2M dataset for molecular adsorption on surfaces, mPFDNN achieves the lowest error metrics (**Figure 3b, c**). This performance advantage is consistently maintained on the large-scale Drug and Alloy dataset, highlighting its broad applicability (**Figure 3d, e**). For the Jarvis benchmark, mPFDNN delivers accuracy in property predictions comparable to that of ALIGNN and outperforms CGCNN models (**Figure 4a**). On the QM9 benchmark for small molecules, mPFDNN yields comparable or superior results across all evaluated properties except for electronic spatial extent ($r^2$) (**Figure 4b**).

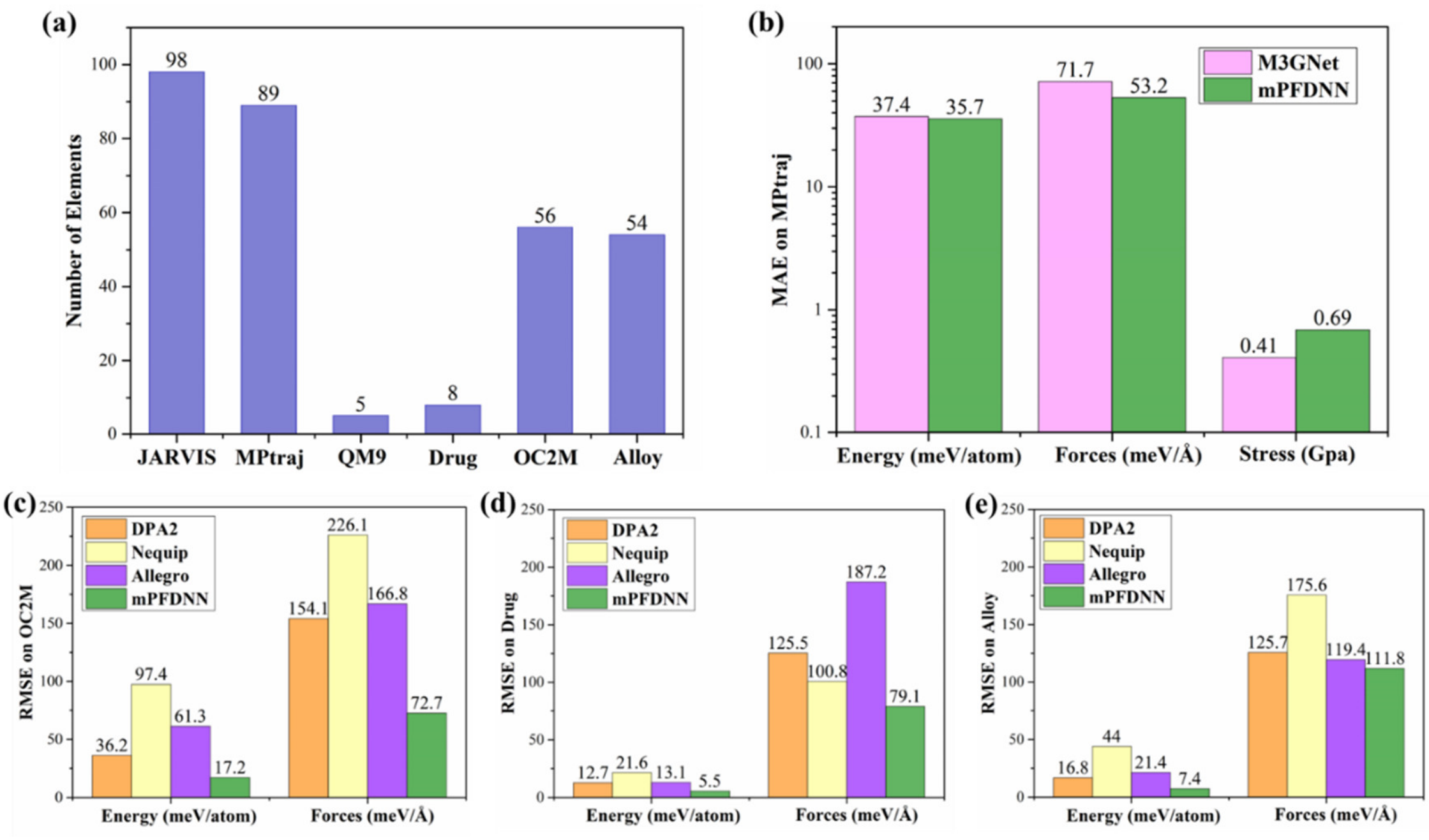

**Figure 3. Benchmark results on force-field datasets. (a)** Elemental coverage of the benchmark datasets. **(b)** The MAE error of dataset Mptraj [26]. **(c)** The RMSE error on dataset OC2M [28]. **(d)** The RMSE error on dataset Drug [27]. (e) The RMSE error on dataset Alloy [29].

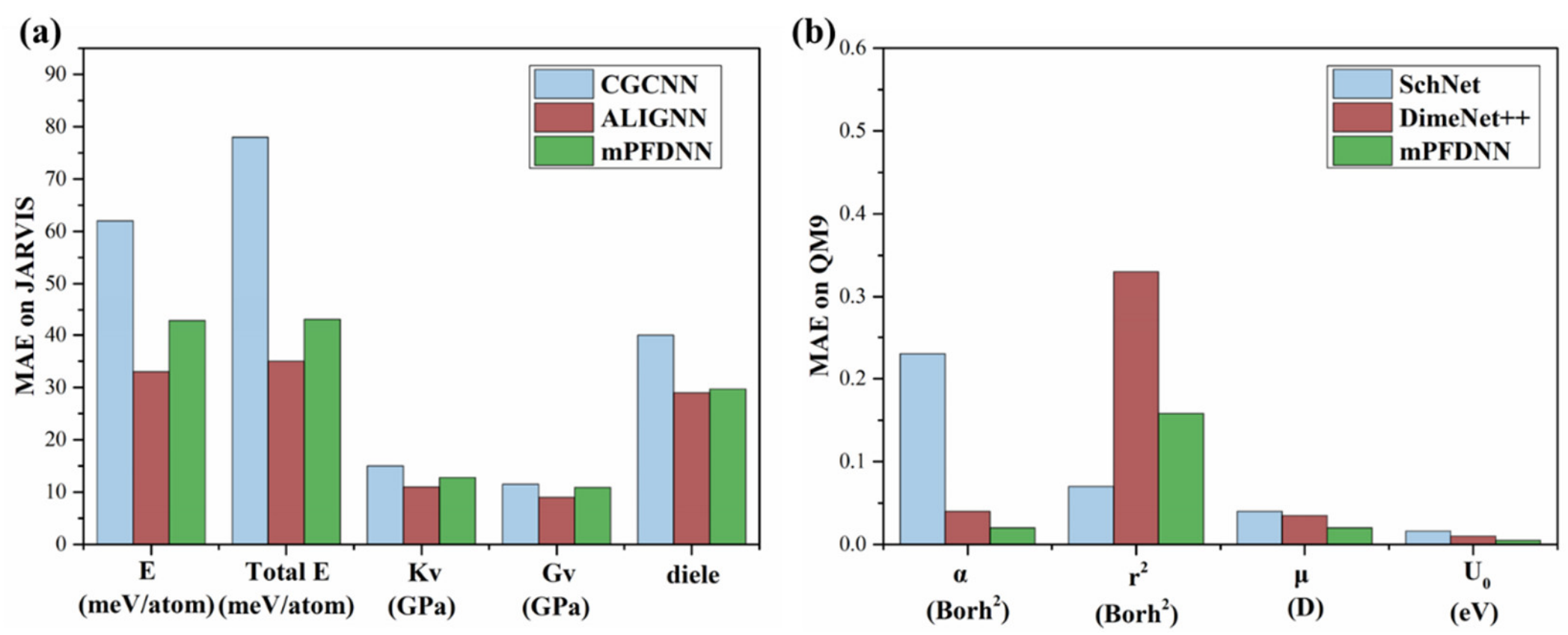

**Figure 4. Benchmark results on property-prediction datasets. (a)** MAE error on dataset JARVIS. **(b)** MAE error on dataset QM9. [25]

Notably, whereas most state-of-the-art models rely on multiple graph-convolution or interaction layers, all mPFDNN models in these benchmarks use only a single recursion step. To explore the effect of recursion depth on model accuracy, we trained five model groups with recursion depths ranging from 0 to 4 using the drug database, which is characterized by significant chemical environmental fluctuations. Our results show that the first recursion yields the most substantial improvement in accuracy, while the model converges after the second recursion (see **Figure S1**). This is because a single recursion qualitatively endows the model with ability to discern atom-specific chemical environments. For example, within the mean-field approximation framework, the model cannot distinguish whether two carbon atoms form a double, single, or π bond. In contrast, the recursively updated representation $\boldsymbol{v}_j^{(t)}$ effectively discriminates each atom's chemical environment, ultimately providing a more accurate description of the interaction.

We further examine the trade-off between model performance and size by varying the channel count ($k$) and orbital truncation ($L'$) of $v_i$ (see **Figure S2**). Our results indicate that increasing $L'$ while maintaining a modest $k$ yields the optimal parameter efficiency. For instance, on the drug dataset, the model with $L'$=2 and $k$=32 achieves comparable accuracy to that with $L'$=0 and $k$=128, yet the parameter count of the former (107,568) is merely one-fifth that of the latter (526,736).

### 3.2 mPFDNN Accurately Captures Ion-Specific Dynamics in Aqueous Solutions

To further evaluate the transferability and physical reliability of mPFDNN, we next considered aqueous electrolyte systems, where subtle long-range non-covalent interactions play a central role. We constructed a dedicated dataset comprising pure water and aqueous solutions (1-4 mol/kg) of NaCl, NaBr, and KCl, totaling 33,819 configurations. The optimized mPFDNN achieved high predictive accuracy (energy MAE: 1.3 meV/atom; force MAE: 36.4 meV/Å), enabling production-level molecular dynamics simulations [30].

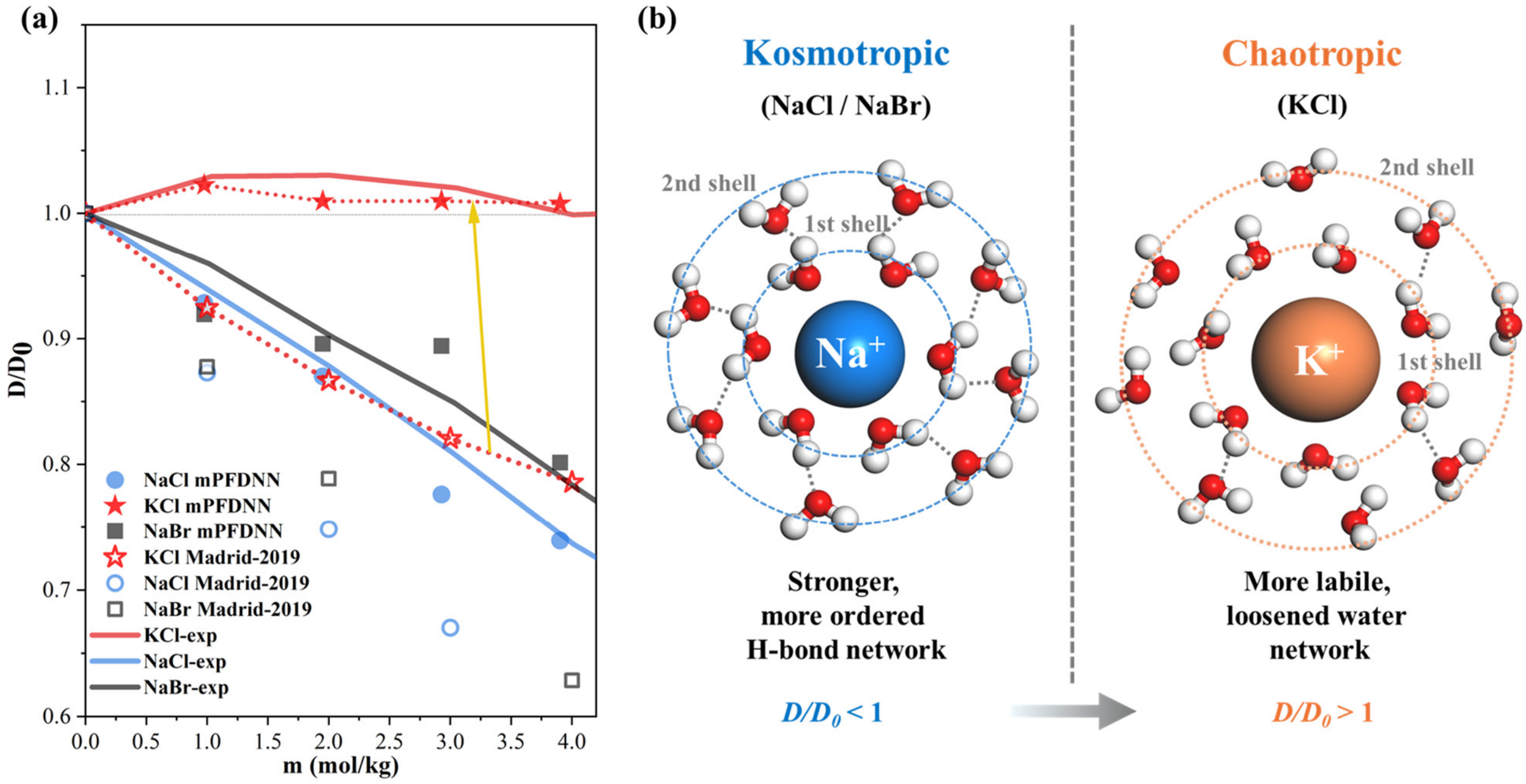

**Figure 5. Ion-specific water diffusion in aqueous salt solutions. (a)** The ratio of the diffusion coefficient of water in NaCl, NaBr, and KCl solutions to that in pure water, as a function of salt concentration calculated by mPFDNN and *Madrid-2019*; **(b)** Schematic comparison of kosmotropic and chaotropic hydration environments, highlighting differences in first- and second-shell organization and their relation to water diffusivity.

A long-standing challenge for classical force fields is capturing ion-specific effects on water dynamics, particularly the counterintuitive enhancement of water diffusivity ($D_{H2O}$) in chaotropic salt solutions (e.g., KCl) [31]. Although the experimental trends associated with the Hofmeister series are well established [32-34], conventional force fields such as *Madrid-2019* [35] fail to distinguish KCl from kosmotropic salts such as NaCl and NaBr, and instead predict reduced water diffusion for all salts [36]. In contrast, mPFDNN-driven molecular dynamics simulations quantitatively reproduce the concentration-dependent $D_{H2O}$ trends for all three salts **(Figure 5a)**, correctly capturing both the reduction for kosmotropes and the enhancement for chaotropic KCl, which is in full agreement with experiments and *ab initio* molecular dynamics (AIMD) benchmarks. Radial distribution function (RDF) analysis revealed the mechanistic basis for this fidelity (**Figure S3**). mPFDNN accurately reproduces the experimental oxygen-oxygen (O-O) RDF of pure water, whereas *Madrid-2019* exhibits clear structural deviations **(Figure S3b)**. Building on this accurate baseline, mPFDNN further reveals pronounced KCl-induced changes in the O-O RDF, especially in the second solvation shell, as supported by the RDF analysis in **Figure S3b** and summarized schematically in **Figure 5b**. These structural rearrangements, which are absent in *Madrid-2019* simulations[35], provide a coherent microscopic explanation for the enhanced water diffusivity uniquely captured by mPFDNN.

By accurately resolving ion-specific hydration effects and their propagation through the hydrogen-bond network, which is a persistent blind spot for classical force fields, mPFDNN demonstrates its suitability for large-scale MLMD simulations of liquid electrolytes and other hydrogen-bonded systems in which accurate ion-solvent cooperativity is essential.

### 3.3 mPFDNN Enables Accurate Adsorption Energy Predictions for HEA Catalysts

High-entropy alloys (HEAs) are a promising class of catalytic materials because their compositional and structural diversity enables tunable reactivity. However, their complexity poses challenges for both experimental screening and computationally intensive ab initio methods [37,38]. Machine learning potentials, particularly the mPFDNN framework validated on the OC2M dataset **(Figure 3c**), provide a transformative pathway by establishing efficient mappings from local site environments to catalytic properties.

A key descriptor of catalytic activity is adsorption energy ($E_{ads}$), which reflects the binding strength of reaction intermediates (e.g., H* in HER, N* and $N_2$* in NRR) to catalytic sites [39-41]. Accurate $E_{ads}$ prediction on HEAs requires capturing both site-specific chemical environments and the accompanying surface relaxations. Traditional DFT methods, which remain computationally expensive for large and chemically complex systems, are not well suited for navigating the vast configurational space of HEAs. Furthermore, static DFT calculations neglect crucial surface structural relaxations, limiting the physical realism of $E_{ads}$ predictions. To address this, we developed a comprehensive workflow integrating mPFDNN with structural optimization. Taking the Ir-Pd-Pt-Rh-Ru HEA (100-atom cell) as a model system , the workflow: (i) constructs unrelaxed surfaces (such as 100, 110, 111, etc.); (ii) relaxes surface slabs using the optimizer embedded with the Alloy mPFDNN model (**Figure 3e**); (iii) identifies adsorption sites (top, fcc, bridge) via a site finder; and (iv) performs adsorption relaxation guided by the optimizer embedded with the HEA mPFDNN model.

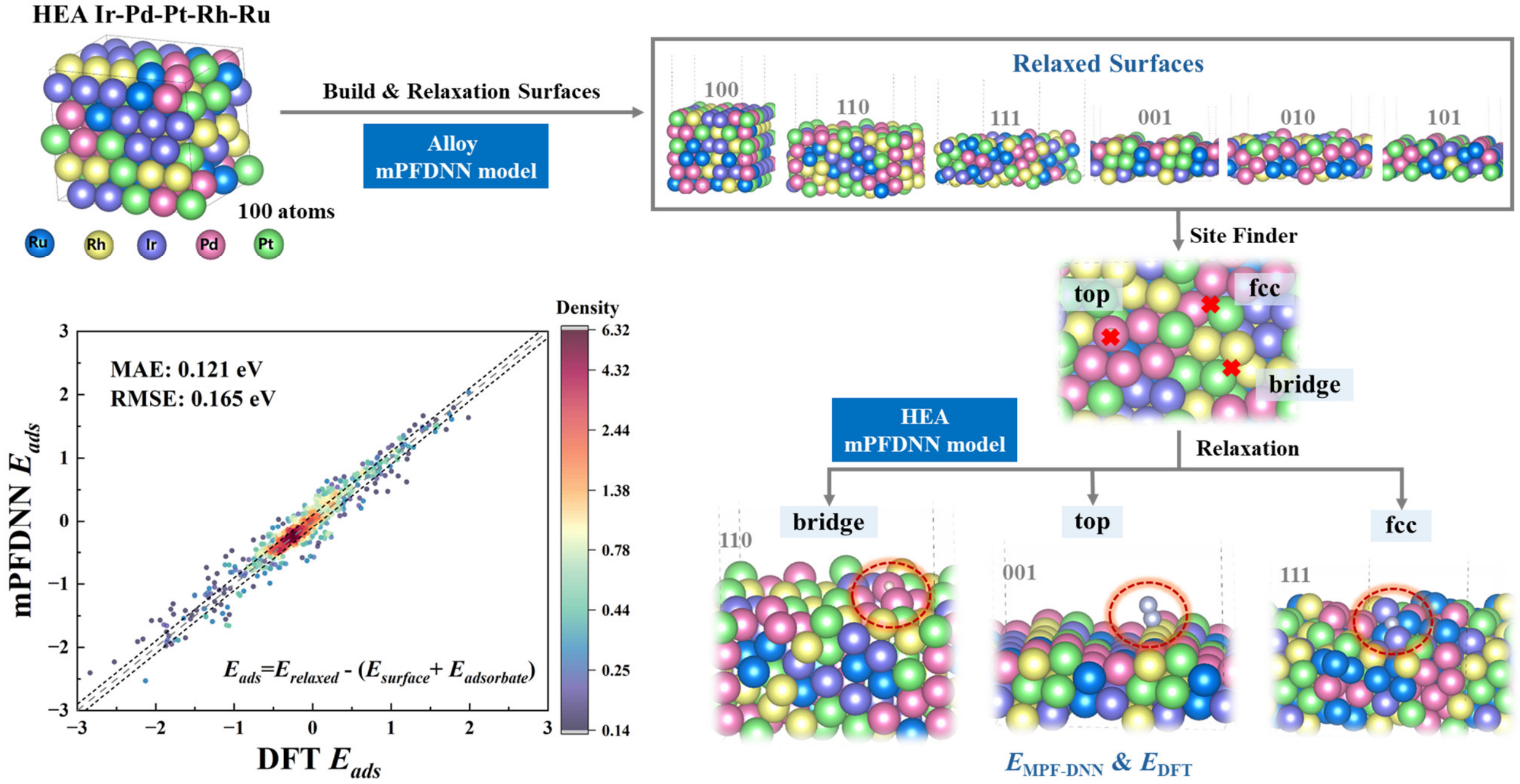

**Figure 6. mPFDNN guided adsorption energy ($E_{ads}$) predictions for HEA Catalysts.** The mPFDNN models are integrated into a workflow that relaxes the high-entropy alloy (HEA) bulk, surface, and adsorption configurations. Guided by an optimizer embedded with the HEA mPFDNN model, this workflow ultimately yields adsorption energy predictions for adsorbates (H, N, and $N_2$).

The optimized mPFDNN model achieved high accuracy (Energy: MAE = 1.3 meV/atom, Force: MAE = 64.7 meV/Å; **Figure S18**) using only 16,772 configurations, merely 5% of the dataset (~355K) required by the state-of-the-art GNN model AGAT [42], which achieves comparable accuracy (1.3 meV/atom, 64.7 meV/Å) on the same HEA system. Notably, mPFDNN achieved this state-of-the-art performance with only 562,448 parameters. This represents a drastic reduction in model size and complexity, approximately 3 to 4 orders of magnitude lower, compared to prevalent large-scale models such as EquiformerV2, which typically contain tens to hundreds of millions of parameters [43,44]. This translates directly to reduced memory footprint, faster inference, and lower training costs, making large-scale HEA exploration computationally feasible. Ultimately, this integrated workflow enables accurate adsorption energy predictions for adsorbates (H, N, and $N_2$), as validated by the correlation between mPFDNN predicted adsorption energies (mPFDNN $E_{ads}$ ) and DFT-calculated ones (DFT $E_{ads}$) shown in the **Figure 6** and **Figure S18**, with metrics like MAE (0.121 eV) and RMSE (0.165 eV) demonstrating the model's efficacy. Notably, unlike the dual-network architecture of the AGAT model, where energy and force are trained separately by two distinct networks, the mPFDNN employs a single unified network with self-consistent energy and forces. Furthermore, mPFDNN demonstrates robust extrapolation to out-of-sample HEA catalysts (**Figure S19**), underscoring its strong generalization capability rooted in its well-founded and parsimonious mathematical formulation.

By combining machine-learned potential with atomic-scale structural relaxation, mPFDNN overcomes the limitations of traditional methods in handling HEA complexity. This approach enables efficient and accurate exploration of HEA configurational space and site-specific catalytic behavior, while maintaining strong extrapolation capability for out-of-domain systems, offering significant potential to accelerate the discovery and rational design of HEA catalysts.

# 4. DISCUSSION

Materials modeling differs from many mainstream artificial-intelligence applications in that it is supported by a substantial body of physical and chemical knowledge about symmetry, locality, and interatomic interactions. An important challenge, therefore, is not only to improve predictive accuracy, but also to incorporate such domain knowledge into model construction in a transparent and scalable manner. In this work, we developed the mPFDNN framework by combining the Material Property Field (MPF) formalism with Hopfield-like recursive dynamics, thereby providing an interaction-based route for structure-property mapping.

The key standpoint of mPFDNN posits that any material properties $\{P\}$ can be formulated as a complicated functional of interatomic interactions, i.e. $\mathcal{F}_P[\varphi_{Ij}(r_{Ij})]$. In this work, we employ the physically-rational form of the MPF $P$ to represent any specific functional $\mathcal{F}_P$. MPF possesses an explicit mathematical formulation, thus distinctions in modeling different material properties are encoded in their corresponding effective pairwise interactions $\tilde{\varphi}_{Ij}$, as $\mathcal{F}_P[\varphi_{Ij}] = (P \circ P^{-1} \circ \mathcal{F}_P)[\varphi_{Ij}] = P[\tilde{\varphi}_{Ij}]$. The construction of $\{\tilde{\varphi}_{Ij}\}$ can be realized in a physically complete and tractable manner. This tractability directly endows mPFDNN with broad applicability across diverse systems and properties, as exhibited in benchmark tests.

Beyond general benchmarks, mPFDNN also shows clear advantages in physically challenging tasks. In Hofmeister-series solutions, our approach captures weak interactions and ion-specific dynamical trends that are difficult to describe with conventional empirical force fields. In the case of adsorption-energy prediction for HEA catalysts, mPFDNN enables efficient exploration of highly heterogeneous configurational spaces while maintaining good extrapolative performance on out-of-domain alloy systems. These examples suggest that the framework is useful not only for standard benchmark prediction, but also for problems in which subtle interactions, chemical complexity, and structural relaxation all play important roles.

In summary, this work demonstrates that interatomic interactions can be incorporated more explicitly into the design of material-oriented neural networks through an analytically motivated recursive framework. By combining physical structure with competitive predictive performance, mPFDNN provides a useful strategy for interaction-aware and physically informed materials machine learning.

# 5. METHODS

## 5.1 Model Training Set-up

All the mPFDNN models employed in this study were implemented with the recursive depth of *t*

= 1, the maximum order of the spherical harmonic expansion $L_{max}$ = 3 and number of channels is set to $k$ = 128. By setting the maximum tensor product order $v$ = 3, the models are capable of capturing 5-body interactions. We adopt spherical Bessel function $j_0^n = \sqrt{\frac{2}{r_c}}\frac{\sin\left(n\pi\frac{r_{Ij}}{r_c}\right)}{r_{Ij}}$ and polynomial truncation functions $f_{cutoff} = 1 - 28\left(\frac{r_{Ij}}{r_c}\right)^6 + 48\left(\frac{r_{Ij}}{r_c}\right)^7 - 21\left(\frac{r_{Ij}}{r_c}\right)^8$ to construct the radial function:

$$R(r_{Ij}) = \sum_{n=1,\ldots,8} j_0^n f_{cutoff} \tag{11}$$

The cutoff radius $r_c$ was set to 6 Å.

The loss function, $\mathcal{L}$, is typically computed as the weighted sum of mean squared errors of target properties $\{P\}$:

$$\mathcal{L} = \sum_{T\in\{\mathcal{T}\}} w_t\,(P - \hat{P})^2 \tag{12}$$

Specifically, in the multi-objective training process of PES, we use the weights for energy, force, and stress $(w_E: w_F: w_S) = (1,10,100)$. All models were trained with the Adam optimizer with $\beta_1 = 0.9$, $\beta_2 = 0.999$, and $\varepsilon = 10^{-8}$ with float32 precision. We use a learning rate of 0.005 and an exponential moving average (EMA) learning scheduler with decaying factor of 0.995. The training process will be terminated when the number of epochs reaches 200, or when no improvement in model performance is observed over 50 consecutive epochs.

### 5.2 Diffusion Simulation

#### 5.2.1 Aqueous Salt Solutions Dataset.

The dataset comprises approximately 33,819 structures of aqueous salt solutions, including water, NaCl, KCl, and NaBr. The dataset was labeled using the SCAN functional, which exhibits significant advantages in describing aqueous systems. Compared to traditional GGA functionals such as PBE (which overestimates hydrogen bond strength) and BLYP (which fails to correct density inversion issues), SCAN, as a meta-GGA functional, accurately balances covalent bonds, hydrogen bonds, and van der Waals interactions. For salt solutions, SCAN also outperforms functionals like revPBE-D3 in describing cation hydration structures. To enhance computational efficiency, the r2SCAN single-point energy calculations were used as initial guesses for the SCAN wavefunction to accelerate convergence.

#### 5.2.2 Aqueous Salt Solutions ML Potential Model

The mPFDNN model employs a three-stage iterative workflow to construct the final network model: (1) initial model training based on the dataset, (2) acquisition of new structures through active learning, and (3) dataset expansion via data labeling using CP2K. During the training, the dataset was partitioned into training, validation, and test sets in an 8:1:1 ratio. Model training was conducted with a learning rate of 0.005, and a maximum of 200 epochs was allowed. Model performance was monitored using the validation set. The final machine learning potential was evaluated on the test set to ensure accuracy, providing a reliable force field for subsequent molecular dynamics simulations. More details of the active-learning process are provided in Section II of the SI.

### 5.2.3 Water Diffusion Coefficient Calculation via MD Simulations

The diffusion coefficients of water ($D_{H2O}$) in salt solutions were determined through MD simulations. Initial structures were generated using PACKMOL [45], constructing systems with 512 water molecules per unit cell and varying salt concentrations (9, 18, 27, and 36 solutes). The simulation box size was determined based on experimental solution density. Structural equilibration was performed using GROMACS with the Madrid-2019 force field, involving 200 ps of NPT and 50 ps of NVT simulations at 300 K to ensure proper density and stability. The molecular dynamics simulations were carried out in LAMMPS [46] using the mPFDNN potential at 330 K (to account for nuclear quantum effects in SCAN [47]) with a 0.5 fs timestep. The simulation included 25 ps of NPT equilibration at 1 bar and 1 ns of NVT process. Water diffusion coefficients were calculated from the NVT trajectory analysis. The diffusion coefficient is calculated through the Mean Squared Displacement (MSD) of the salt solution. MSD is defined as the average of the squared displacement of water molecules.

$$\mathrm{MSD}(t) = \langle |r(t) - r(0)|^2 \rangle \tag{13}$$

Where r(t) is the position of a water molecule at time t, and ⟨⟩ represents the average over all water molecules.

The diffusion coefficient D is directly related to the slope of the MSD:

$$D = \lim_{t \to \infty} \frac{1}{6t} \mathrm{MSD}(t) \tag{14}$$

According to the Einstein relation for three-dimensional diffusion, the water diffusion coefficient was obtained from the slope of the MSD.

The trajectory from the 1 ns NVT production simulation was utilized for this calculation. The MSD was obtained using the built-in msd command in LAMMPS, and the diffusion coefficient D was derived from the MSD using the Einstein relation formula. Each diffusion coefficient was determined from five independent simulations, and the final value was averaged from the diffusion coefficients obtained from the simulation trajectories, see SI Section III.

## 5.3 HEA Dataset and Adsorption Energy Calculation Methods

The dataset consists of approximately 16,772 configurations of Ir–Pd–Pt–Rh–Ru high-entropy alloy (HEA) surface slabs (with H, N, and $N_2$ adsorbates) and bulk structures. All structures in the dataset was labeled by the *Vienna Ab Initio Simulation Package* (VASP) [48,49] with periodic boundary conditions and the projector augmented wave (PAW) pseudopotentials [50]. To ensure consistency and complementarity with the OC20 dataset, we adopted identical computational parameters [28]. The plane-wave basis set was employed with a kinetic energy cutoff of 350 eV. Exchange and correlation effects were treated within the generalized gradient approximation (GGA) using the revised Perdue-Burke-Ernzerhof (RPBE) functional [51,52], which provides an improved description of the energetics for atomic and molecular adsorption on surfaces. Bulk and surface calculations were performed considering a K-point mesh for the Brillouin zone derived from the unit cell parameters as an on-the-spot method, employing the MonkhorstPack grid [53].

The adsorption energy ($E_{ads}$) for HEA systems was calculated according to the following expression: $E_{ads} = E_{relaxed} - (E_{surface} + E_{adsorbate})$, where $E_{relaxed}$ is the total energy of the relaxed adsorbate-substrate system, $E_{surface}$ is the energy of the clean, relaxed HEA surface, and $E_{adsorbate}$ is the energy of the free adsorbate in the gas phase. A negative $E_{ads}$ value corresponds to a thermodynamically favorable adsorption process.

## ASSOCIATED CONTENT

**Supporting Information:** See Supporting Information for: (i) the recursion of $v_I^{(t)}$ in model dynamics, (ii) aqueous solution ML potential modeling and dataset specifications, (iii) diffusion simulation convergence, and (iv) mPFDNN performance for in-domain and out-of-domain HEA systems.

**Data availability:** All related datasets and model files in this work can be found in https://doi.org/10.5281/zenodo.17717505 [26].

**Code availability:** The source code for mPFDNN is available on *GitHub*: https://github.com/yecaichao/mPFDNN.

## AUTHOR INFORMATION

**Author contributions:** Y.H., Y.S., C.Y. and W.Z. conceived the initial idea. C. Y., W.Q. and X.X. constructed datasets. Y.H. and Y.S. developed and formalized the code base. W.Q. and C.Y. performed the simulations and analyzed the data. W.Z., C.Y., Y.W., J.Y. and W.A.G offered insight and guidance throughout the project. Y.H., Y.S., C.Y. and W.Z wrote the manuscript and contributed to the discussion and revision.

## ACKNOWLEDGMENTS

This project is supported by the National Natural Science Foundation of China (92463310, 92163212, 52473235), National Key R&D Program of China (2022YFA1203400), Guangdong Basic and Applied Basic Research Foundation (2025A1515011494), Guangdong Provincial Key Laboratory of Computational Science and Material Design (2019B030301001), Shenzhen Science and Technology Program (JCYJ20250604144327038) and High Level of Special Funds (G03050K002). Computing resources were supported by the Center for Computational Science and Engineering at Southern University of Science and Technology.

**Notes**

The authors declare no competing interests.

**Support Information for**

# Material-Property-Field-based Deep Neural Network in Hopfield Framework

**Table of Contents:**

### Section I: The Recursion of $v_I^{(t)}$ in Model Dynamics

In our model, the Lagrangian function for hidden neurons $L_h$ is defined to be the property field $\mathcal{T}$, which can be reformulated as:

$$\mathcal{T} = \sum_I \mathcal{T}_I = \sum_I \prod_{j \neq I} (\varphi_{Ij}(\boldsymbol{r}_{Ij}, Z_I, Z_j) + 1) \tag{S1}$$

$$= \sum_I \exp(\sum_j \ln(\varphi_{Ij} + 1))$$

$$= \sum_I \exp\left(\sum_j \sum_\beta (\frac{(-1)^\beta}{\beta} (\varphi_{Ij})^\beta)\right)$$

Here, we define $\sum_j (\frac{(-1)^\beta}{\beta} (\varphi_{Ij})^\beta)$ as the hidden neurons $h_\mu$ with the composite index $\mu = (I, \beta)$, thus $L_h = \sum_I \exp(\sum_\beta h_{(I,\beta)})$. Notably, under this definition,

$$\frac{\partial L_h}{\partial h_\mu} = \frac{\partial \sum_{I'} \exp(\sum_{\beta'} h_{(I',\beta')})}{\partial h_{(I,\beta)}} = \exp(\sum_{\beta'} h_{(I',\beta')}) \equiv P_I \tag{S2}$$

$P_I$ is $\mu$-independent, so the update of feature neurons follows $v_I^{(t+1)} = P_I^{(t)}$, as $\sum_\mu^{N_h} \xi_{I\mu} = 1$.

A more general recursive process in mPFDNN can be written in the following form:

$$\boldsymbol{A}_I^{(t+1)\mathbf{LM}} = \sum_j \sum_{\substack{\alpha \\ \in(\boldsymbol{lmL'M'}) \to \mathbf{LM}}} \boldsymbol{c}_\alpha \boldsymbol{C}_{\boldsymbol{lmL'M'}}^{\mathbf{LM}} [\boldsymbol{v}_{jk}^{(\mathrm{t})\boldsymbol{L'M'}} \otimes \mathfrak{R}_{\boldsymbol{lm}}(\boldsymbol{r}_{Ij})] \tag{S3}$$

This constitutes a sophisticated tensor product process, wherein α denotes the specific pathway for reducing the tensor product into irreducible representations. **Figure S1** illustrates the model convergency on drug dataset with the deep of recursion.

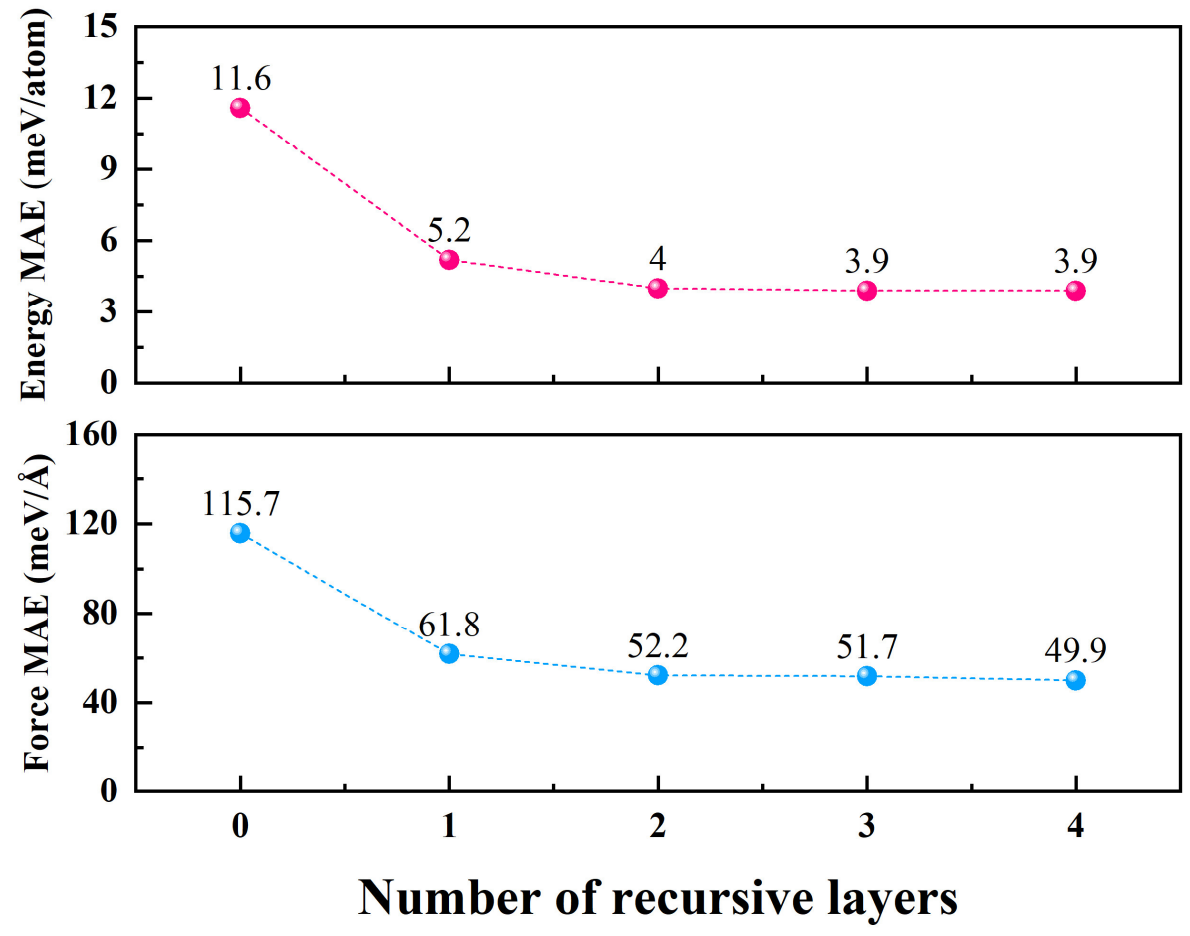

**Figure S1.** The convergence of model accuracy on drug dataset with the number of recursions.

Our framework incorporates a truncation of the orbital angular momentum $L'$ (of $v_{jk}$) to manage computational expense. Within this setup, we identify that appropriately increasing $L'$ is a more parameter-efficient strategy for boosting accuracy than expanding the channel count $k$, as shown in **Figure S2**. Besides, to prevent excessive model complexity, the maximum order of $\boldsymbol{L}$ is truncated at $\boldsymbol{l_{max}}$.

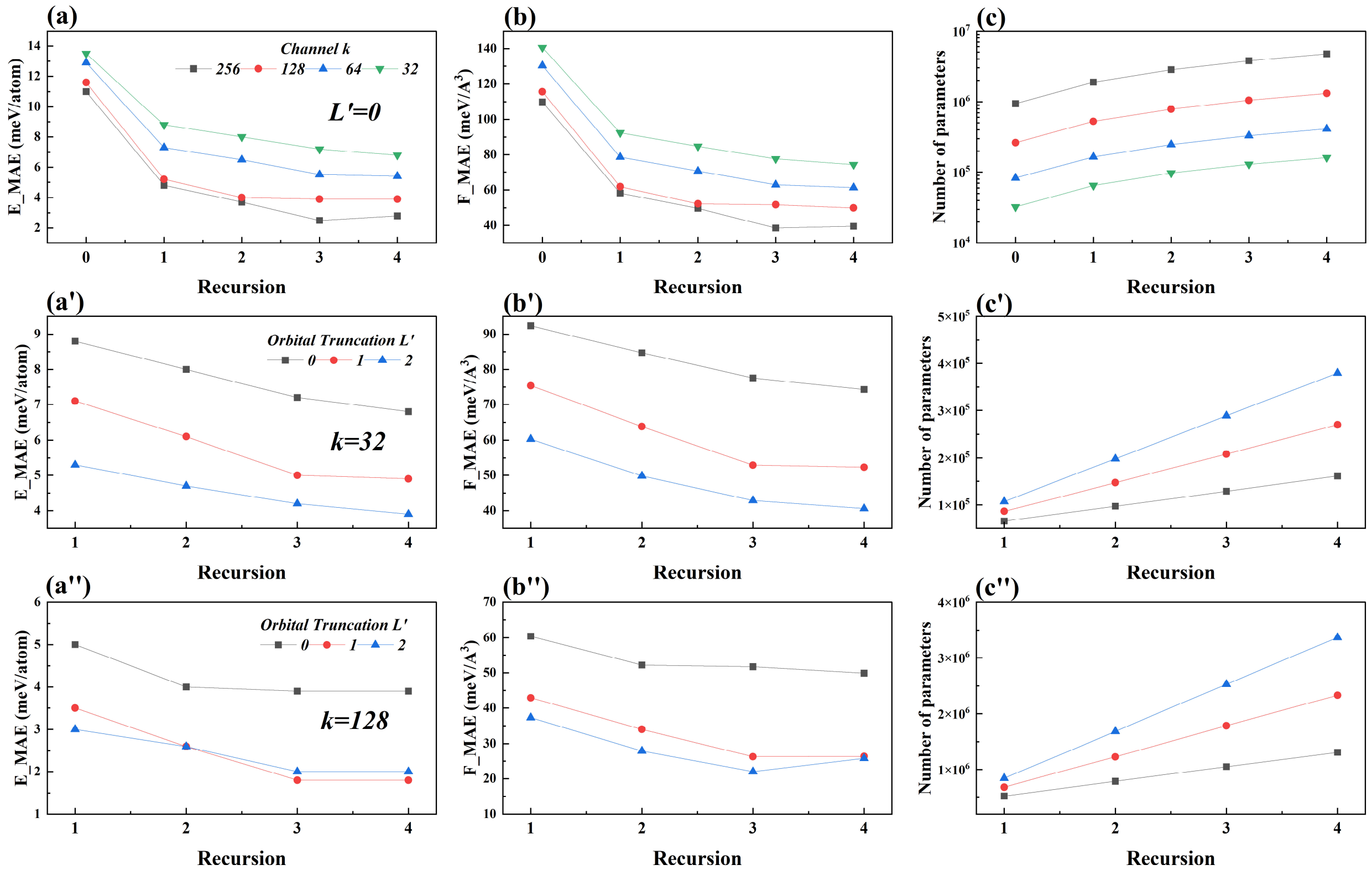

**Figure S2.** Dependence of model performance and size on recursive depth under different truncation and channel conditions: (a-c) fixed $L'$=0, (a'-c') fixed $k$=32 and (a''-c'') fixed $k$=128.

## Section II: Aqueous Solution ML Potential Modeling and Dataset Specifications.

The mPFDNN model employs a three-stage iterative workflow to construct the final network model: (1) initial model training based on the dataset, (2) acquisition of new structures through active learning, and (3) dataset expansion via data labeling using CP2K. During the training, the dataset was partitioned into training, validation, and test sets in an 8:1:1 ratio. Model training was conducted with a learning rate of 0.005, and a maximum of 100 gradient propagation steps was allowed. Model performance was monitored using the validation set. The final machine learning potential was evaluated on the test set to ensure accuracy, providing a reliable force field for subsequent molecular dynamics simulations. In the active learning process, the mPFDNN method was used to extract atomic local environment descriptors as structural features. These descriptors were analyzed using the k-means clustering algorithm: (1) the cluster centers for each descriptor category were computed, (2) the frequency of each category across the dataset was calculated, and (3) a frequency score was generated based on the descriptor distribution and frequency, serving as a quantitative measure of structural rarity. The system prioritized the 20% of data with the lowest frequency scores as initial structures for molecular dynamics simulations, ensuring that the active learning process focused on rare configurations. Iterations were terminated when the frequency scores of newly selected structures reached 80% of the dataset, indicating comprehensive coverage of the target conformational space.

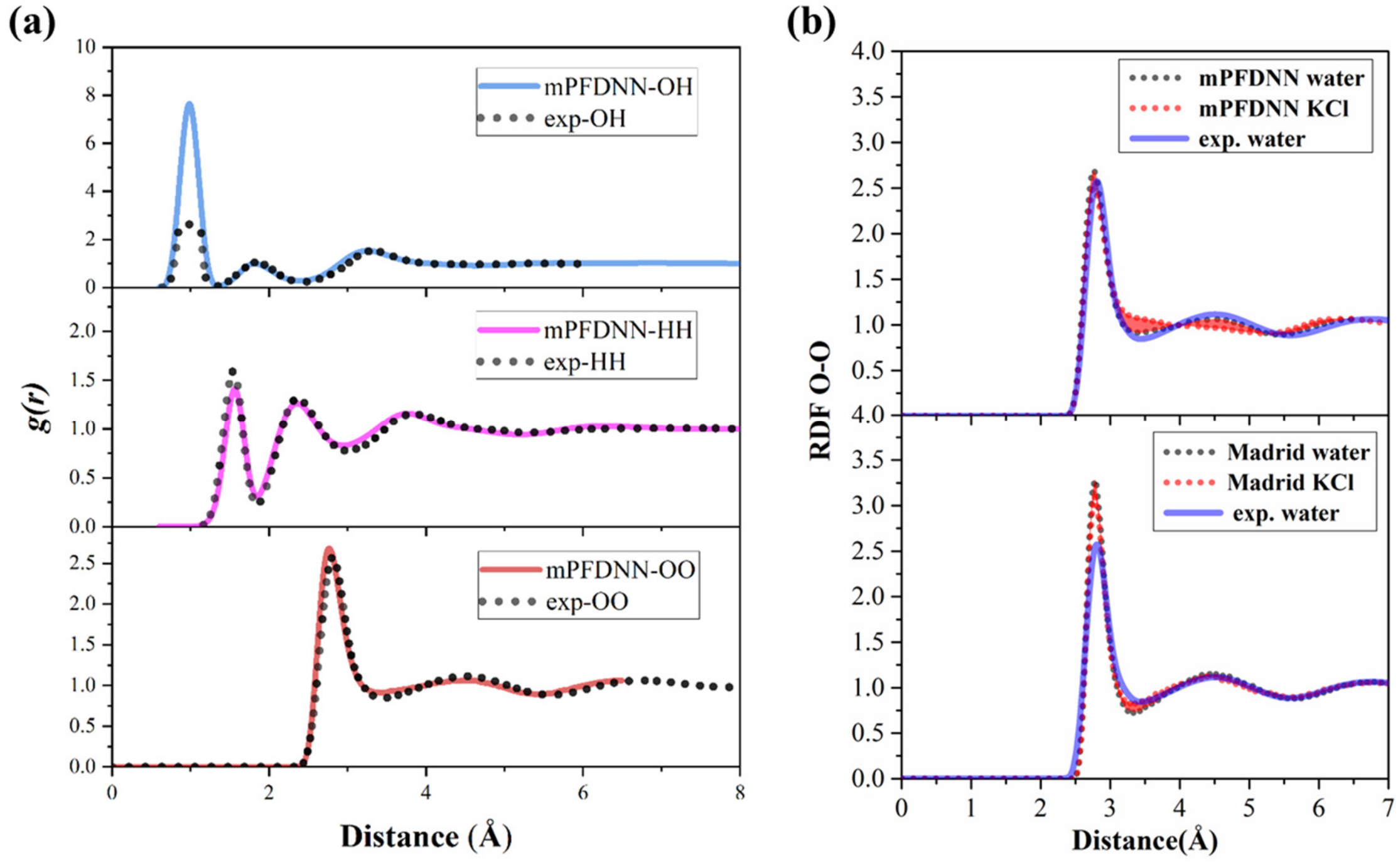

**Figure S3.** (a) Radial distribution function (RDF) of water simulated by mPFDNN model compared to the experiments; (b) RDF of water (O-O) in water and KCl solutions calculated by mPFDNN and *Madrid-2019*.

**Section III: Diffusion Simulation Convergence.**

The following is an approximate process for calculating the diffusion coefficient:

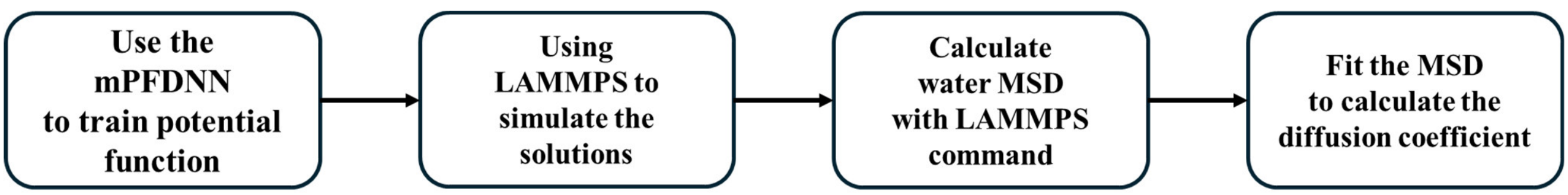

**Figure S4.** Process for calculating the diffusion coefficient

The diffusion coefficients of the salt solutions were obtained using molecular dynamics simulations. The initial structures for the molecular dynamic simulations were generated by the PACKMOL software, where different concentrations of salt solutions were constructed by adding 9, 18, 27, and 36 salt solutes into 512 water molecules. When generating the initial structures, the box size was calculated based on the experimental solution density; subsequently, the initial structures were preliminarily adjusted using the Madrid-2019 empirical force field through GROMACS software: 200 ps of NPT ensemble simulation and 50 ps of NVT ensemble simulation were performed at 300 K to ensure density and structural stability. The molecular dynamics simulations for the analysis part were performed using LAMMPS software, driven by the mPFDNN network model. The SCAN functional typically employs molecular dynamics simulations at 330 K to compensate for the absence of nuclear quantum effects at 300 K, so all simulations were conducted at 330 K, including 25 ps of NPT ensemble equilibration and 1 ns of NVT ensemble production simulation, with a uniform time step of 0.5 fs. The structures were all under periodic boundary conditions, and the pressure during the NPT stage was set to 1 bar. The diffusion coefficients were obtained by analyzing the trajectories from the NVT production simulations.

The diffusion coefficient is calculated through the Mean Squared Displacement (MSD) of the salt solution. MSD is defined as the average of the squared displacement of water molecules.

$$\mathrm{MSD}(t) = \langle |r(t) - r(0)|^2 \rangle \tag{S8}$$

Where r(t) is the position of a water molecule at time t, and ⟨⟩ represents the average over all water molecules.

$$D = \lim_{t\to\infty} \frac{1}{6t} \mathrm{MSD}(t) \tag{S9}$$

The diffusion coefficient D is directly related to the slope of the MSD.

Based on the Einstein relation applicable to three-dimensional diffusion systems, the water diffusion coefficient in salt solutions was calculated.

The trajectory from the 1 ns NVT production simulation was utilized for this calculation. The MSD was obtained using the built-in compute msd command in LAMMPS, with msd output every 100 steps, and the diffusion coefficient D was derived from the MSD using the Einstein relation formula. Each diffusion coefficient is determined by five independent simulations, each with the same initial structure and random initial velocities. The final value is obtained by averaging the diffusion coefficients derived from the simulation trajectories.

The diffusion coefficient ($D$) calculation part is derived from the fitting of the mean square displacement (MSD) curve. Below are the MSD curves for each system, where the blue curve represents MSD, and the grey curve represents the diffusion coefficient calculated after fitting the MSD.

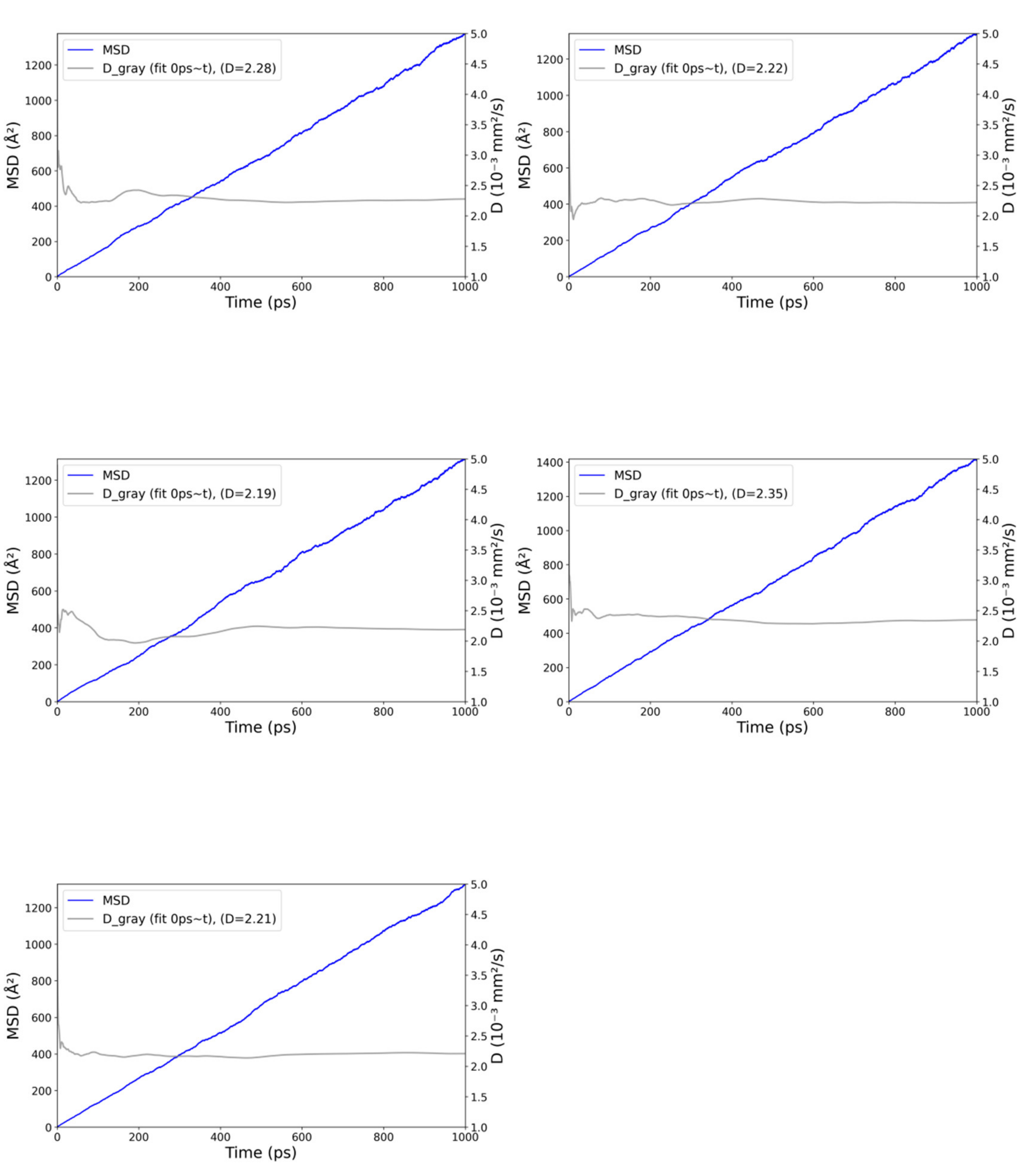

**Figure S5.** The MSD curve of pure water

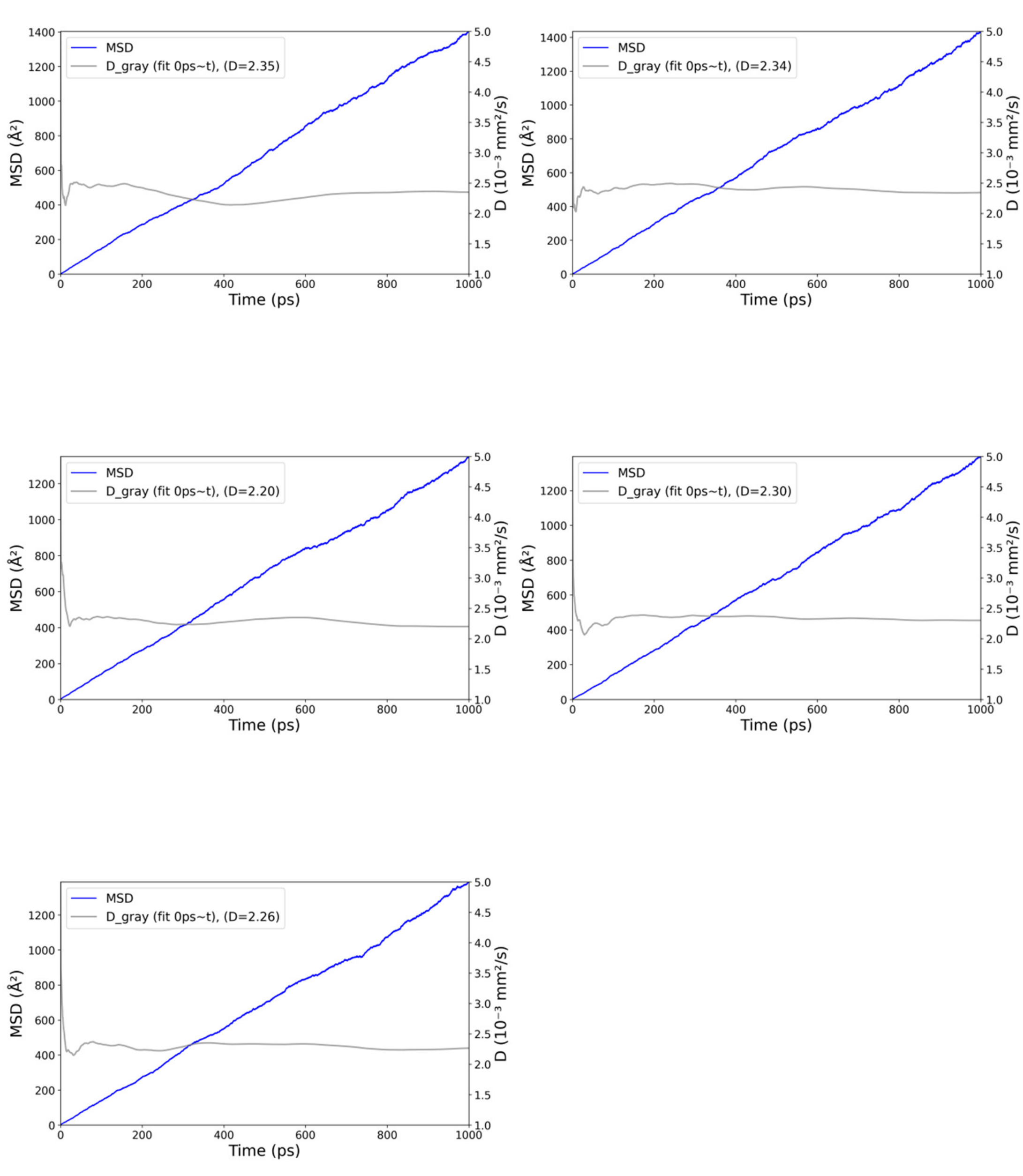

**Figure S6.** The MSD curve of potassium chloride (KCl - 0.9757 mol/kg)

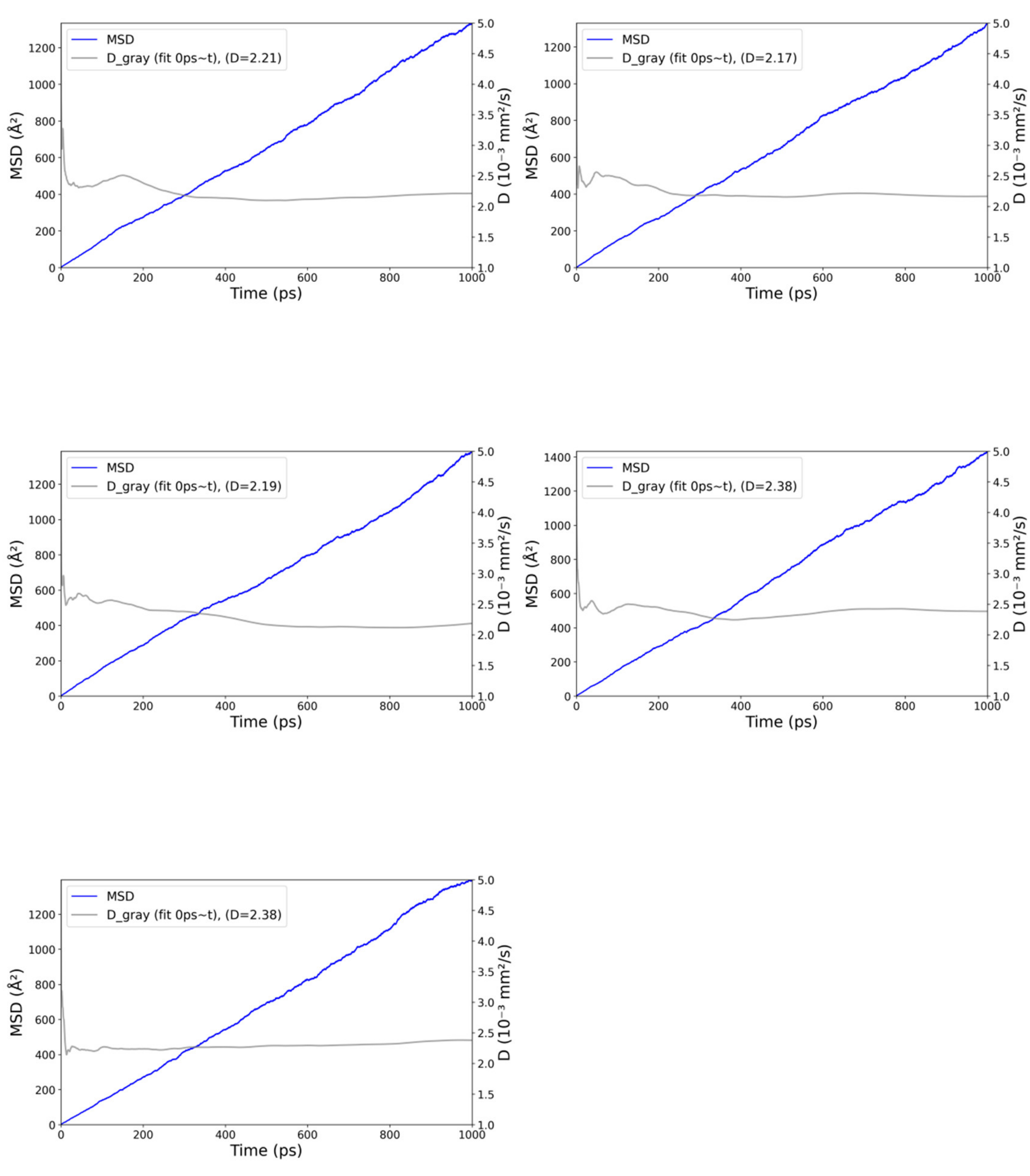

**Figure S7.** The MSD curve of potassium chloride (KCl - 1.9515 mol/kg)

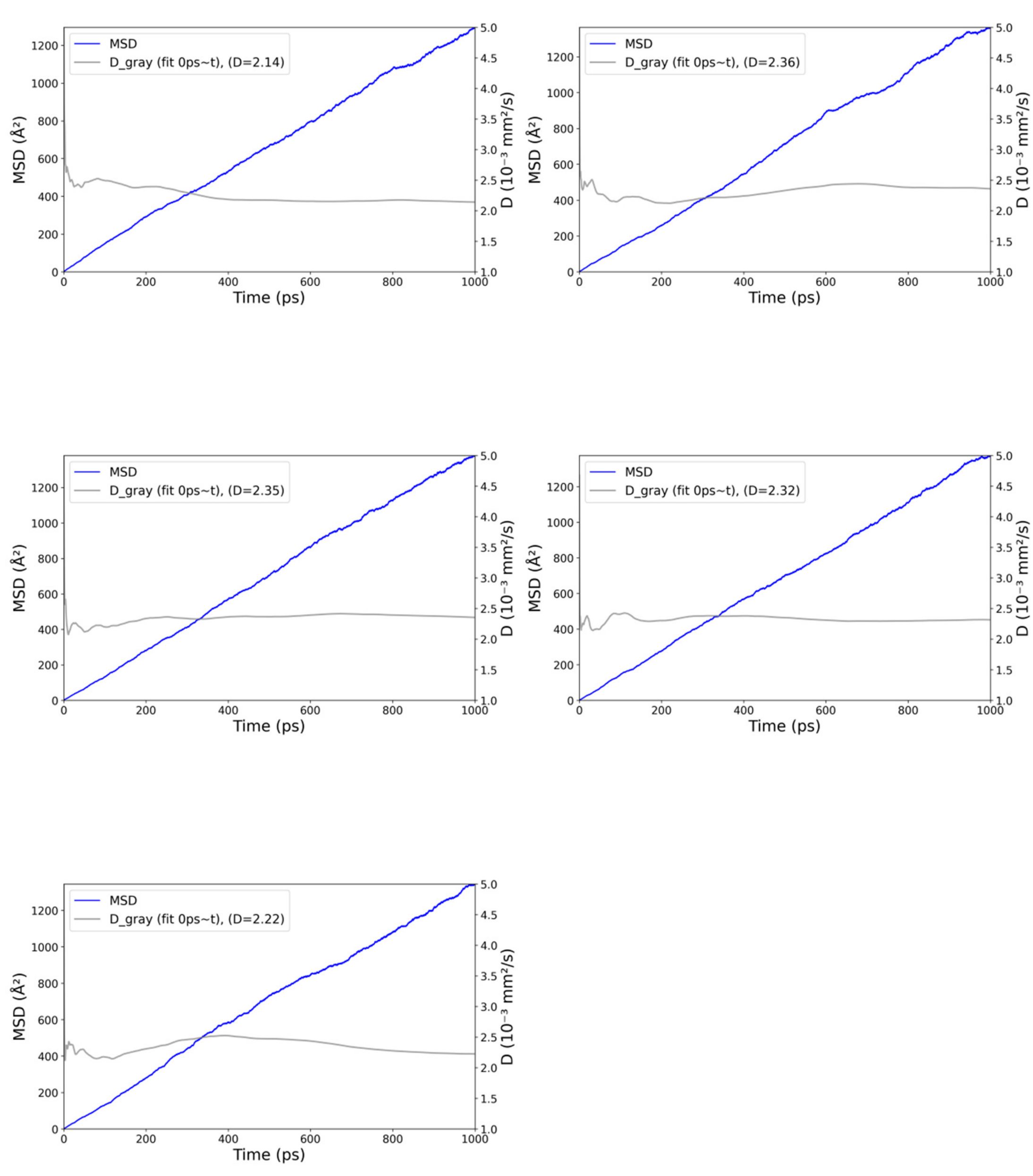

**Figure S8.** The MSD curve of potassium chloride (KCl - 2.9272 mol/kg)

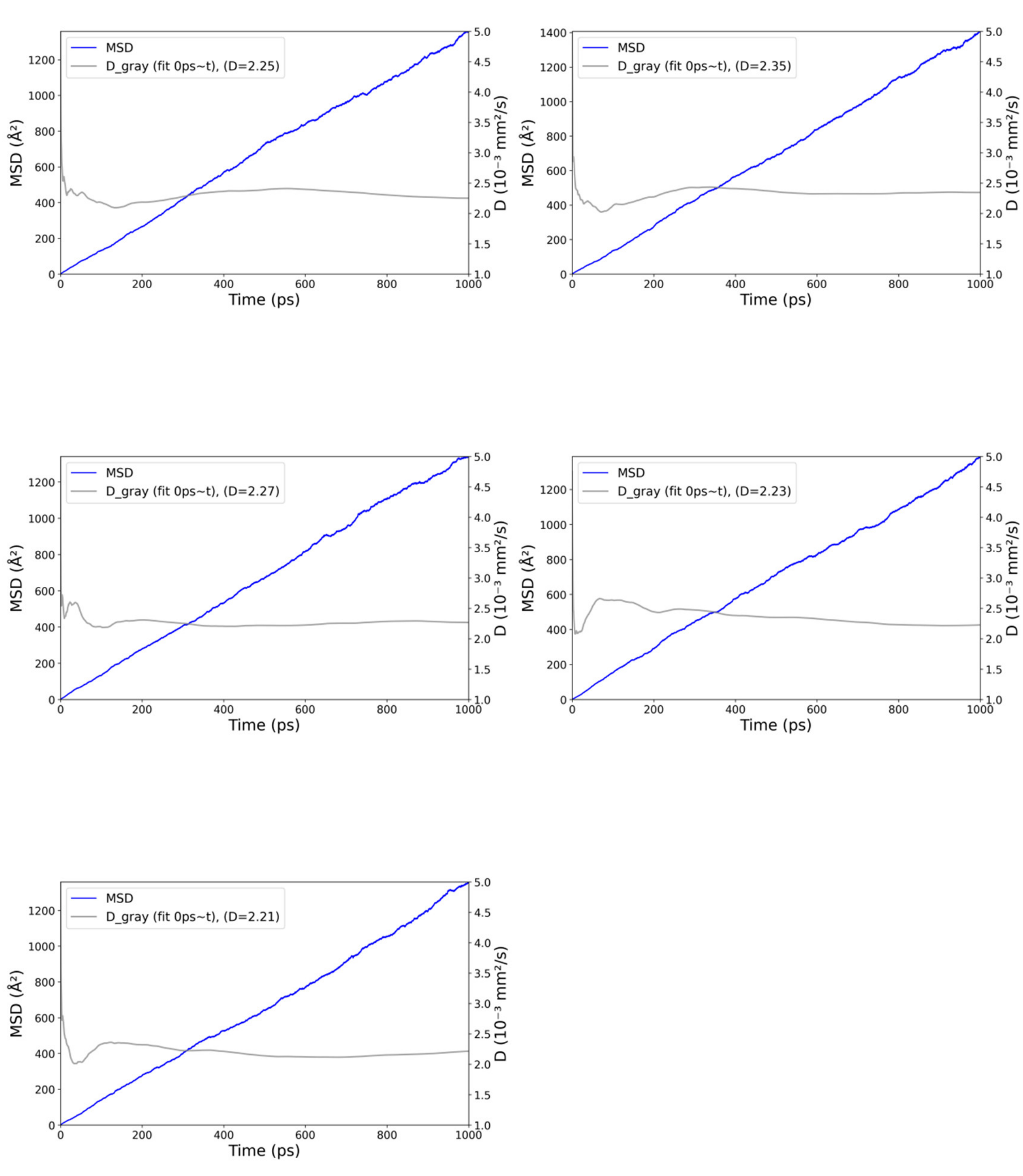

**Figure S9.** The MSD curve of potassium chloride (KCl - 3.9029 mol/kg)

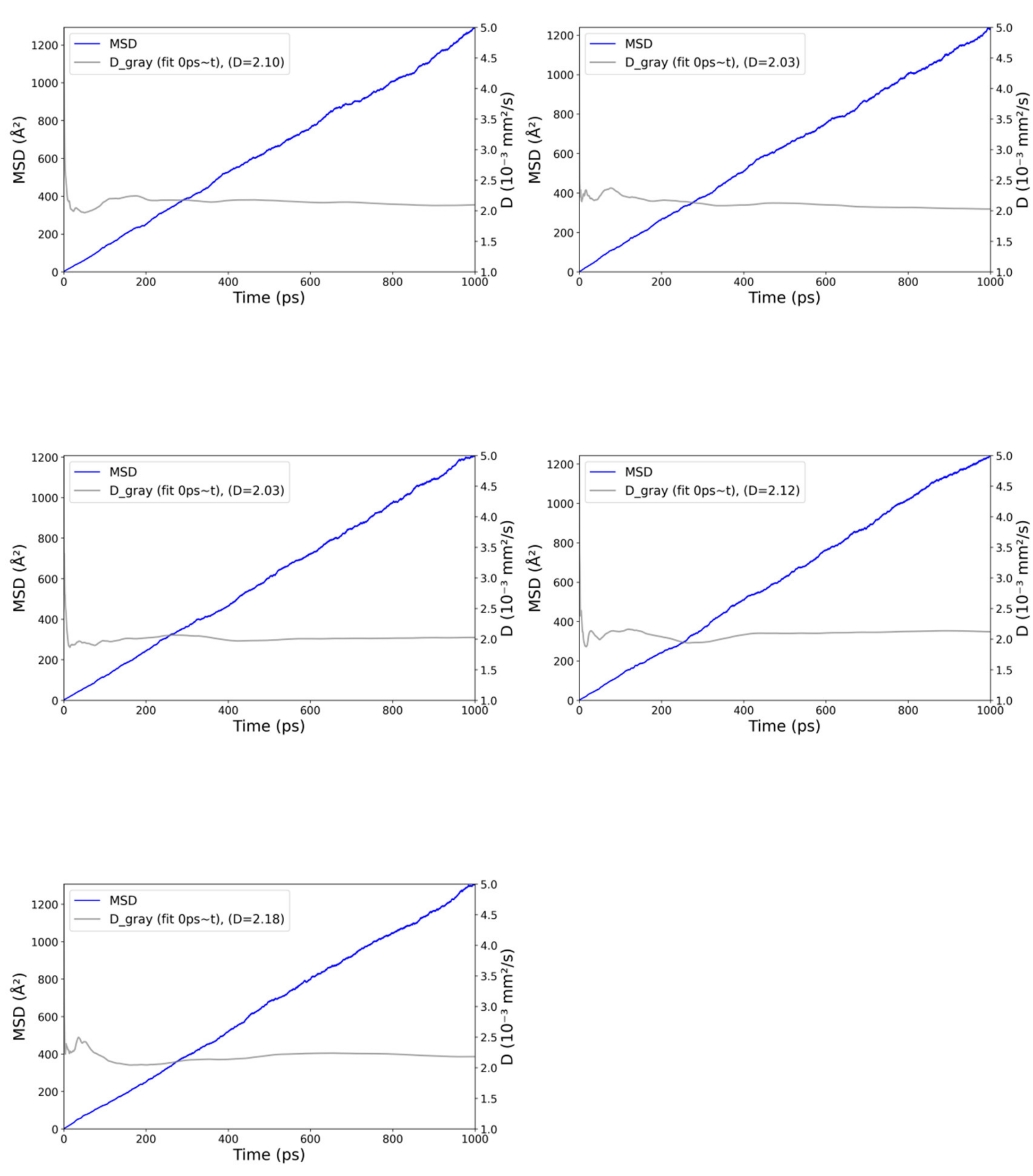

**Figure S10.** The MSD curve of sodium chloride (NaCl - 0.9757 mol/kg)

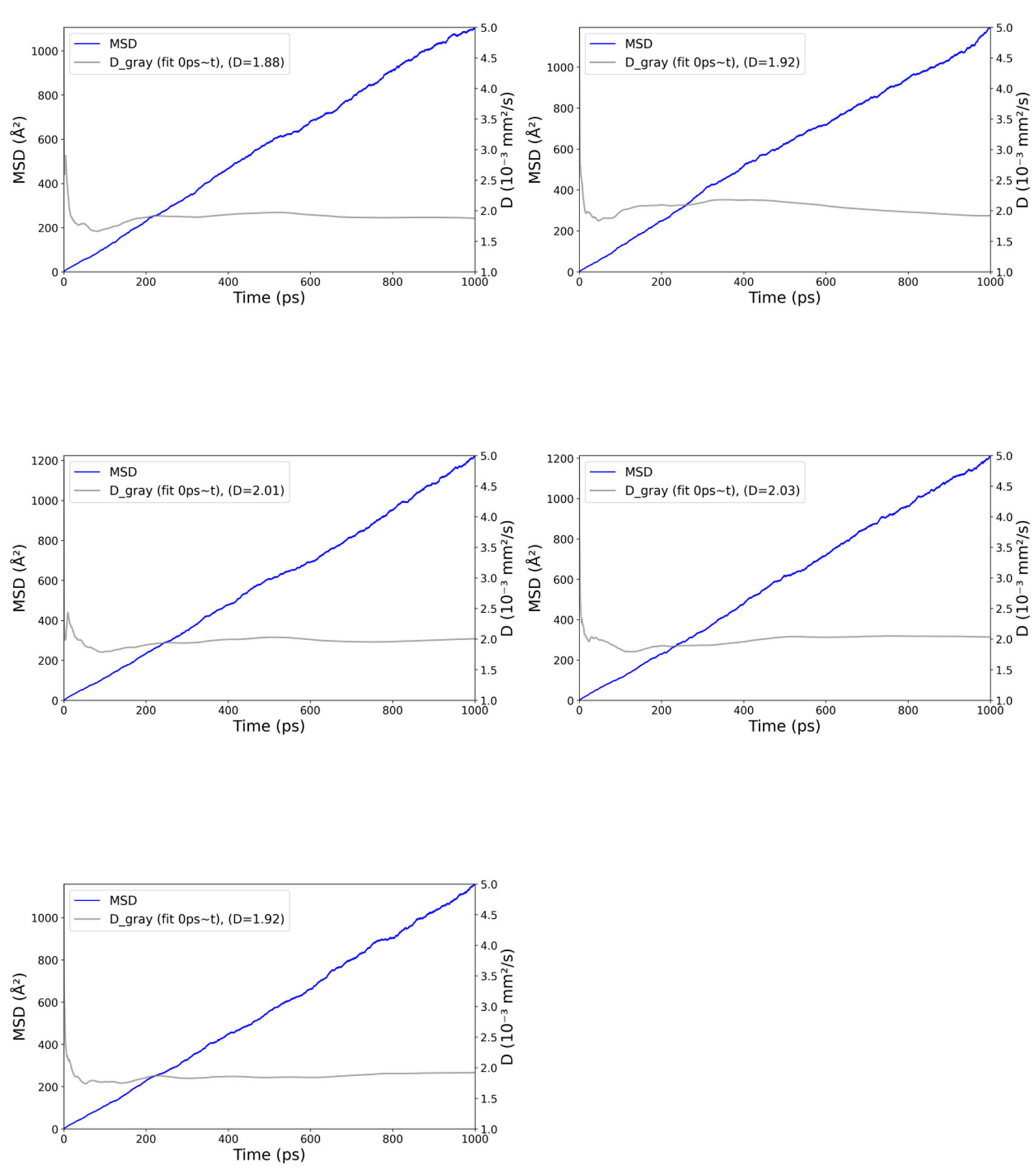

**Figure S11.** The MSD curve of sodium chloride (NaCl - 1.9515 mol/kg)

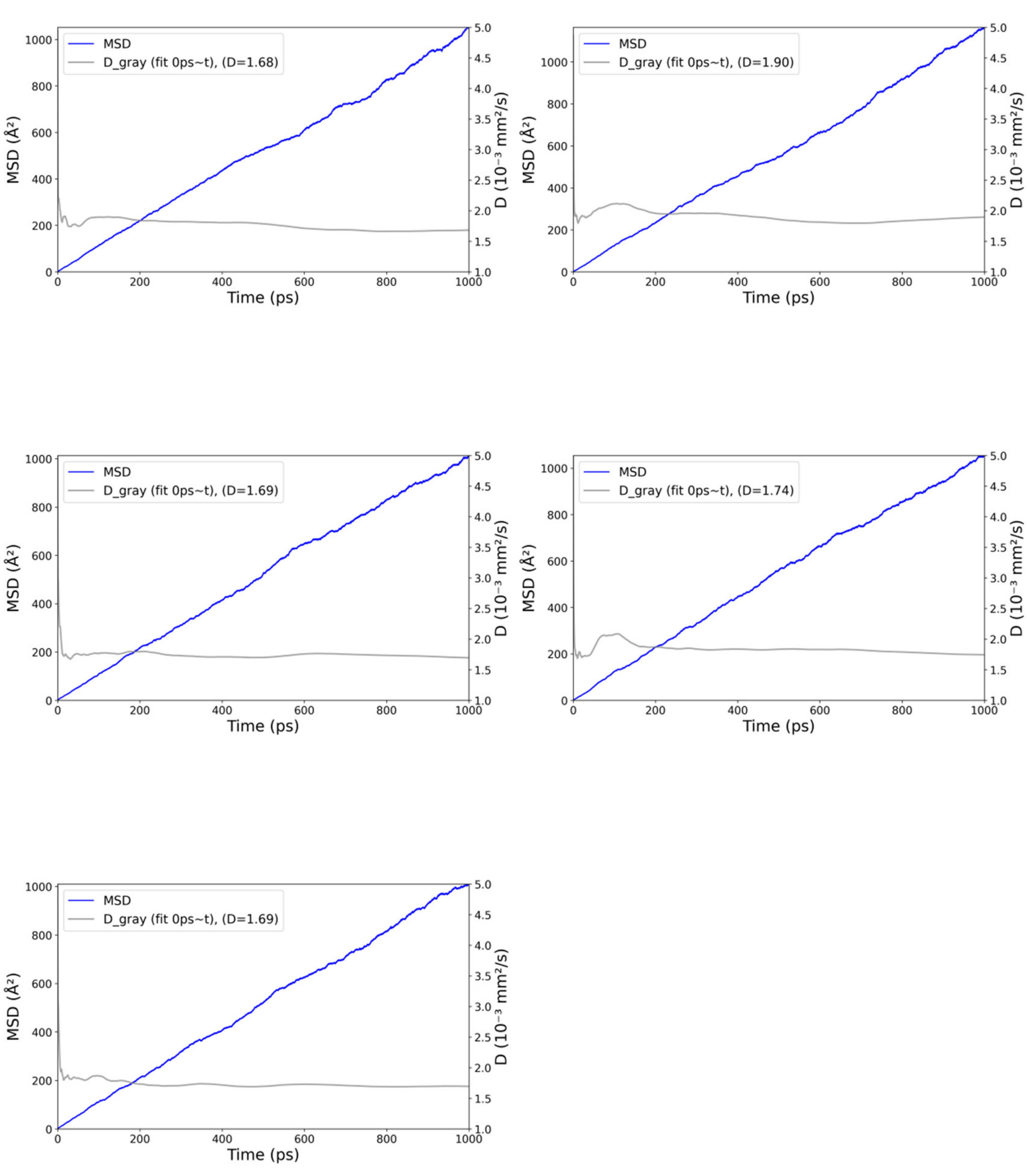

**Figure S12.** The MSD curve of sodium chloride (NaCl - 2.9272 mol/kg)

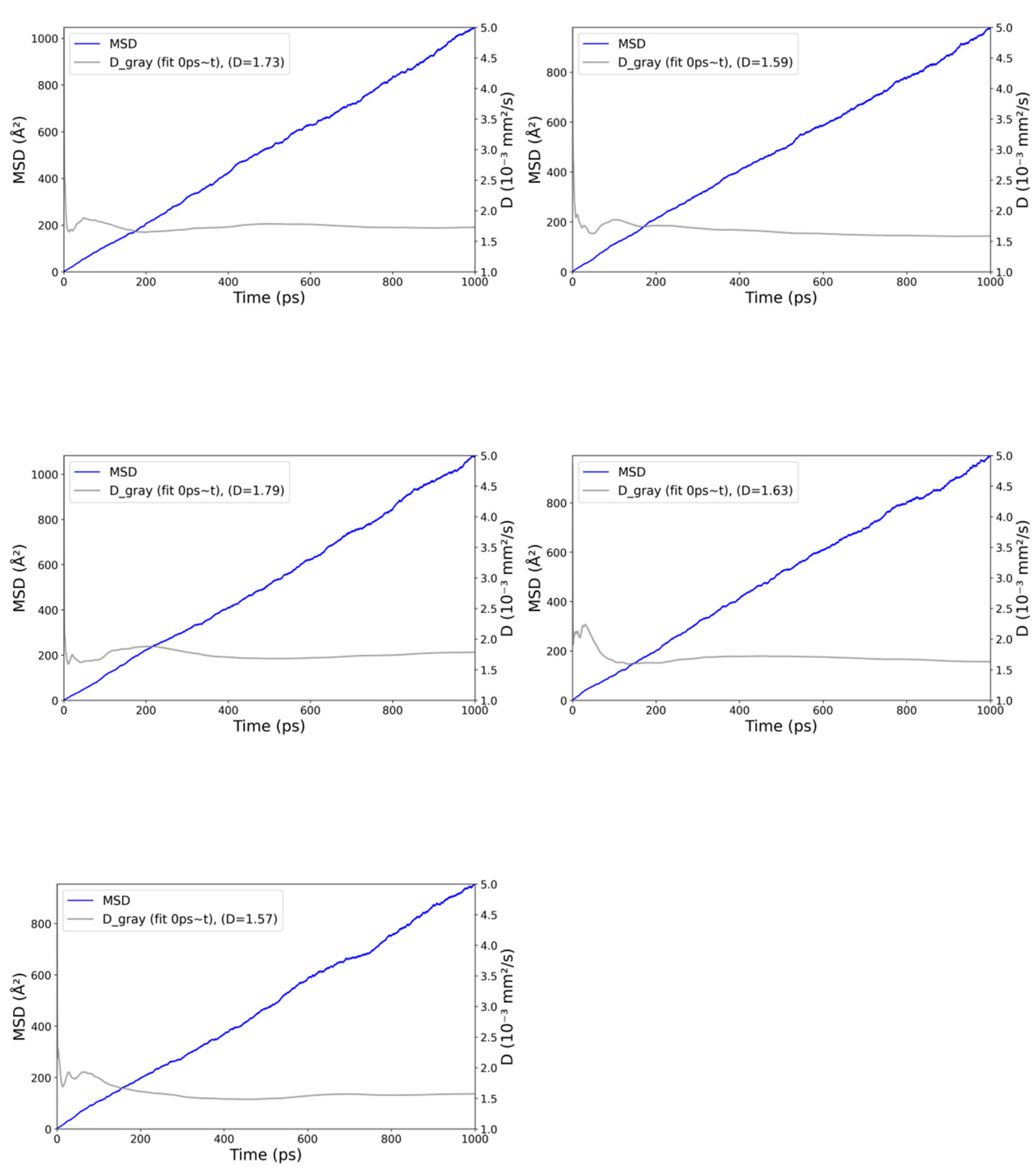

**Figure S13.** The MSD curve of sodium chloride (NaCl - 3.9029 mol/kg)

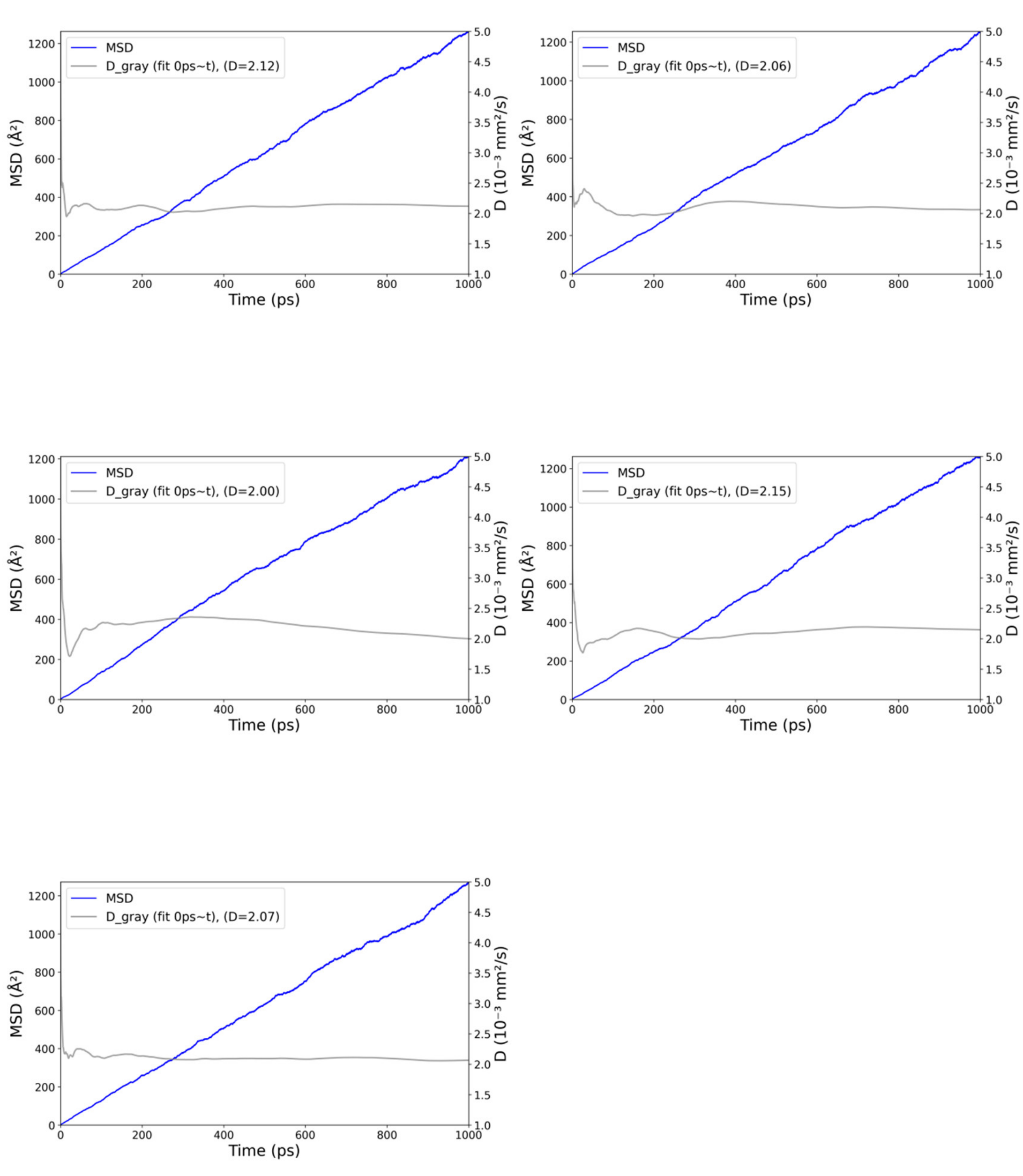

**Figure S14.** The MSD curve of sodium bromide (NaBr - 0.9757 mol/kg)

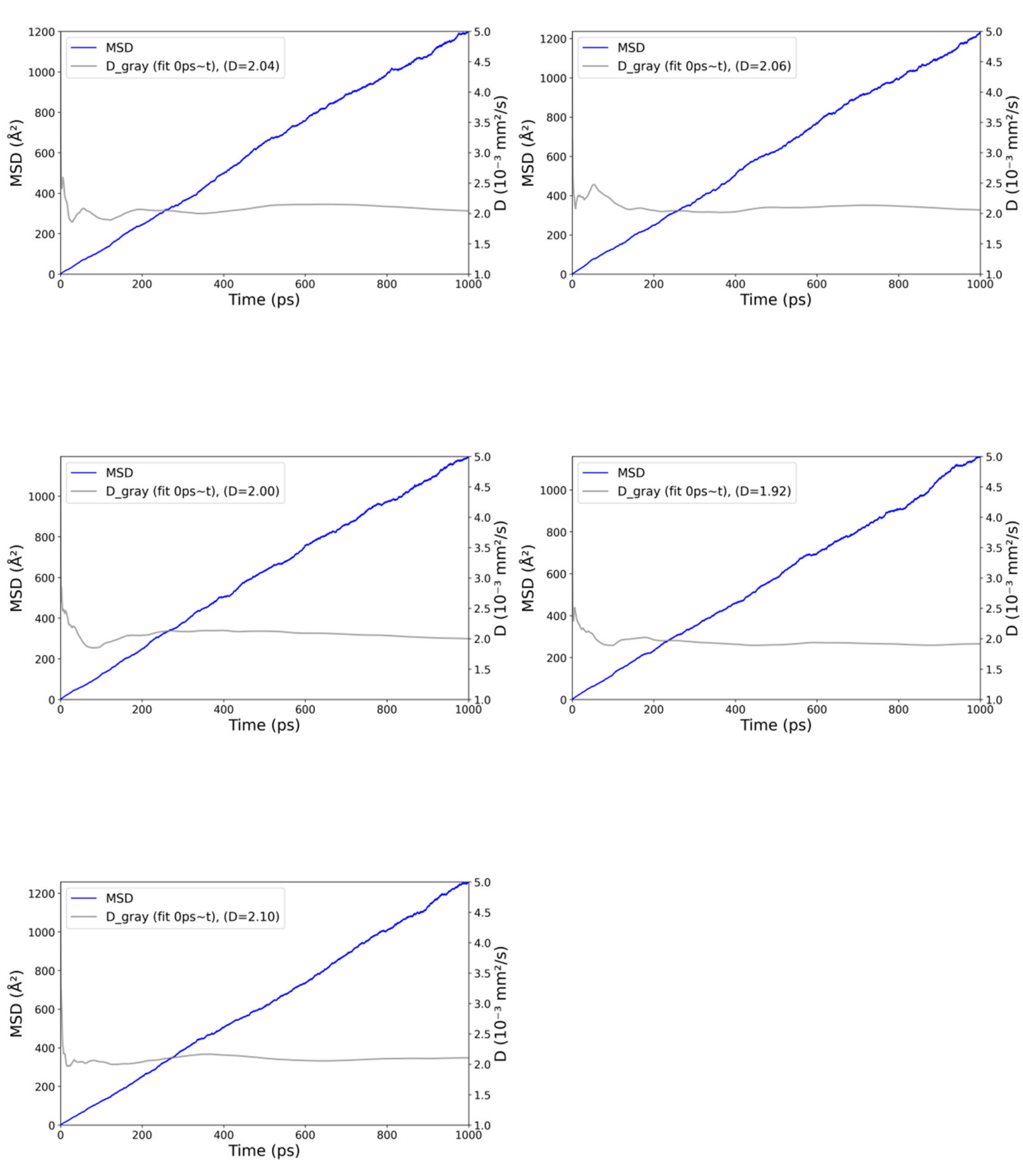

**Figure S15.** The MSD curve of sodium bromide (NaBr - 1.9515 mol/kg)

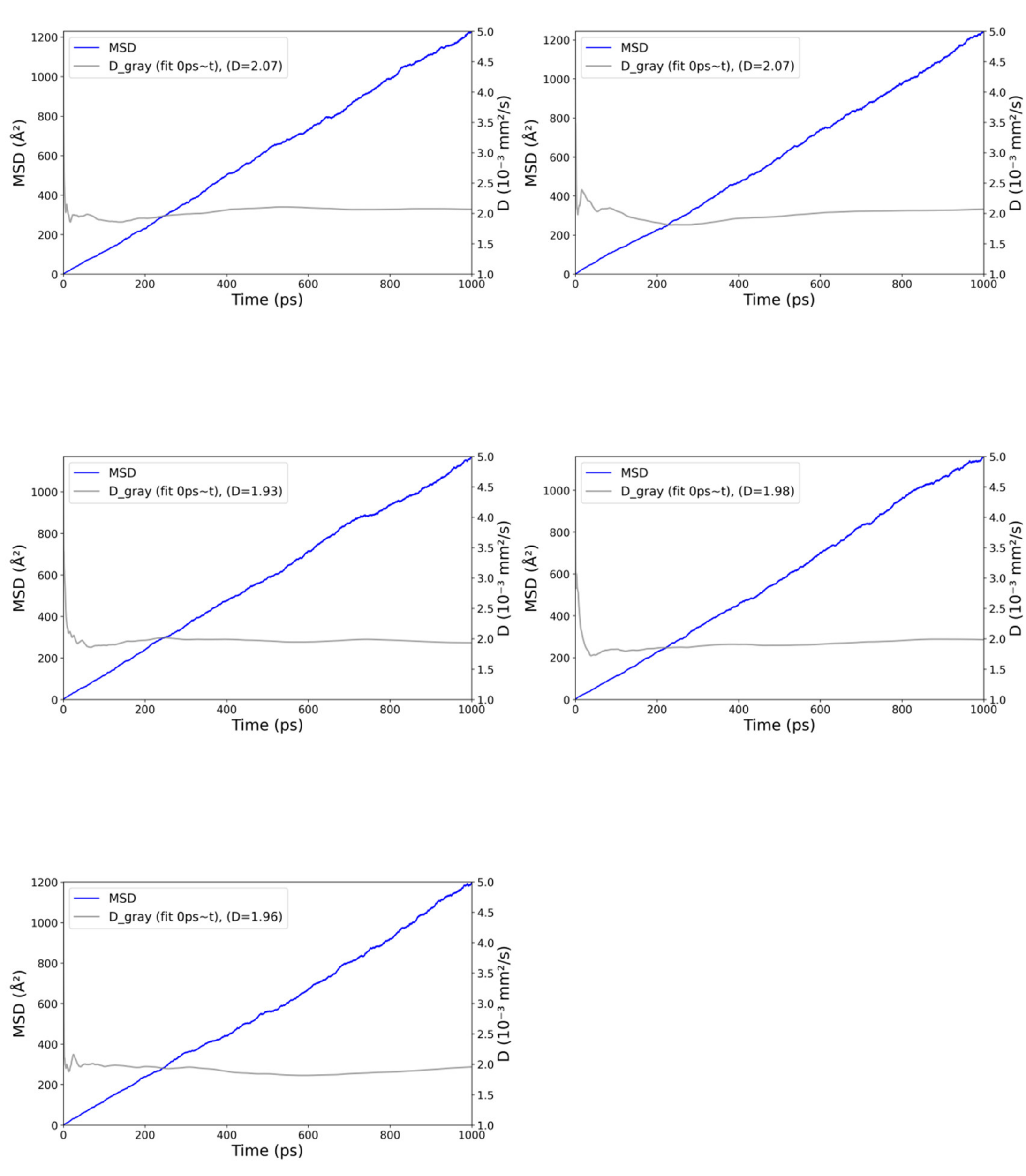

**Figure S16.** The MSD curve of sodium bromide (NaBr - 2.9272 mol/kg)

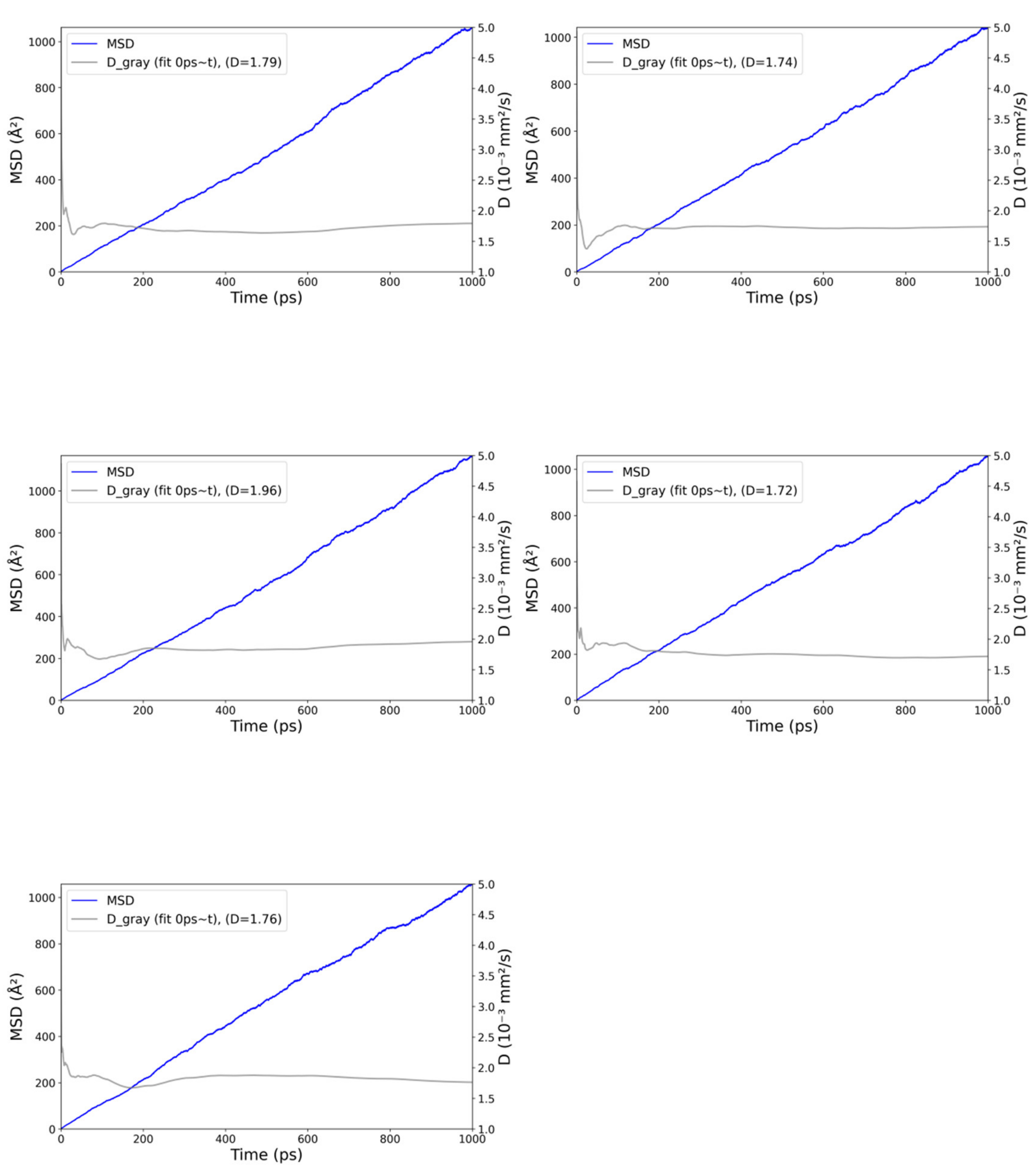

**Figure S17.** The MSD curve of sodium bromide (NaBr -3.9029 mol/kg)

## Section IV: mPFDNN Performance for In-domain and Out-of-domain HEA Systems.

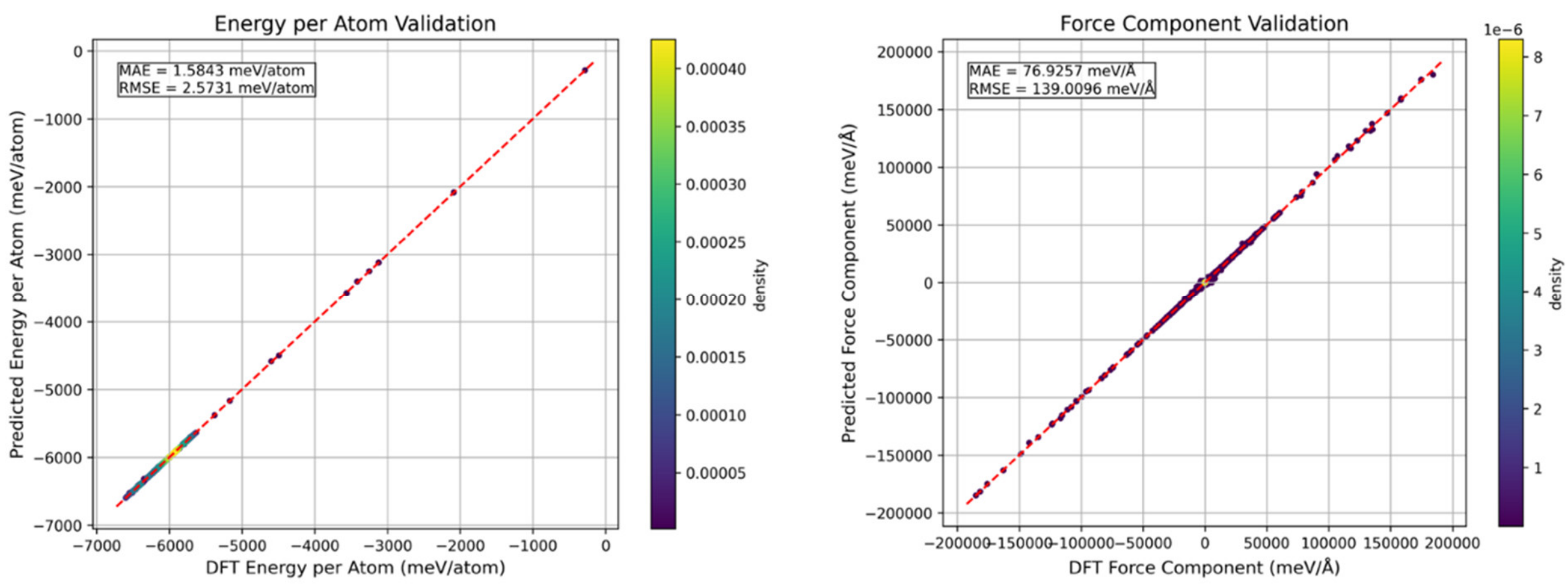

**Figure S18.** mPFDNN performance in energy and force prediction for HEA catalysts

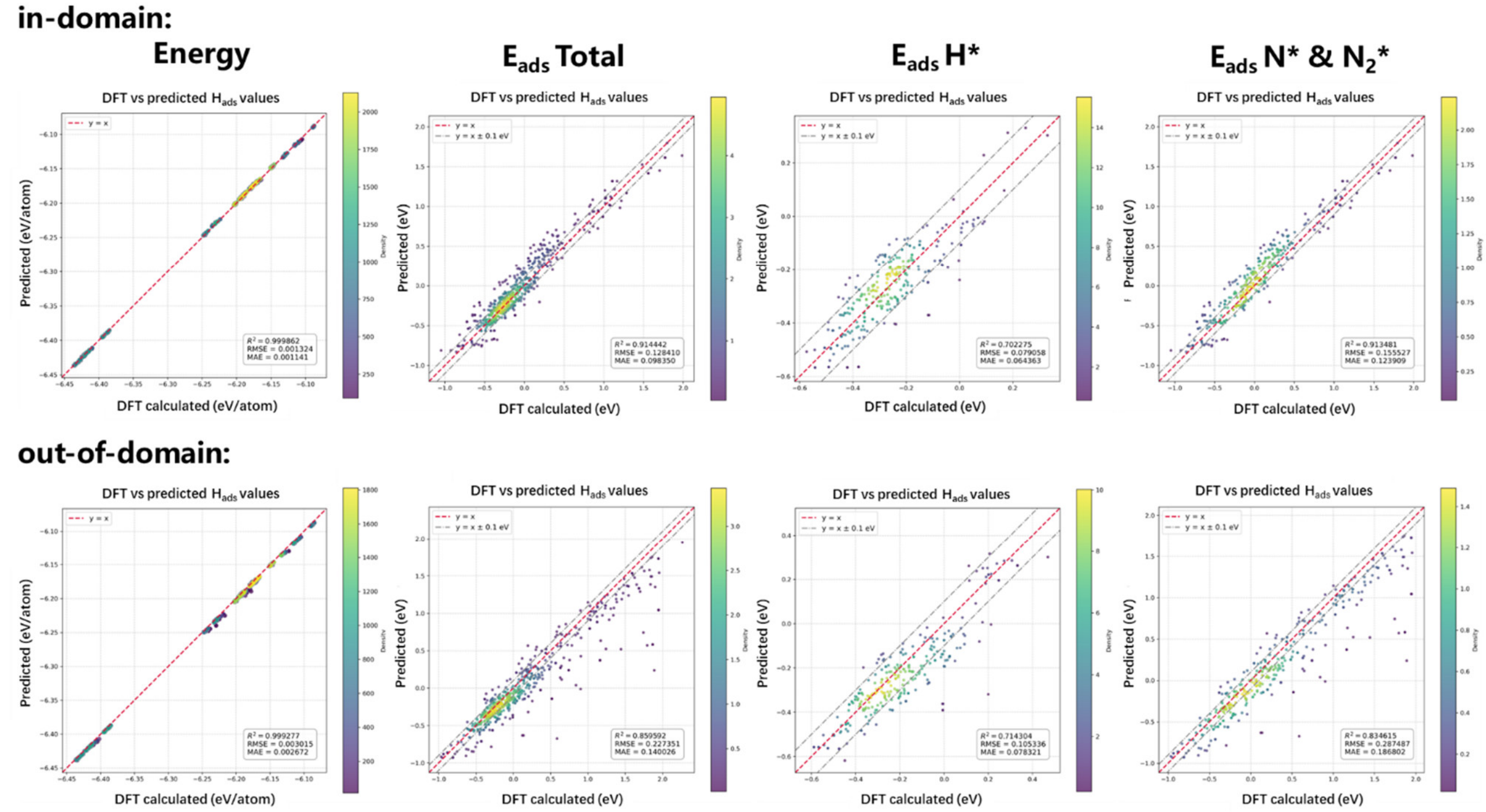

**Figure S19.** mPFDNN performance in adsorption energy prediction for out-of-sample HEA catalysts